# Measuring the Sizes, Shapes, Surface Features and Rotations of Solar System Objects with Interferometry

Jian-Yang Li[a,*], Marc J. Kuchner[b], Ronald J. Allen[c], Scott S. Sheppard[d]

[a]Department of Astronomy, University of Maryland, College Park MD, 20742, USA

[b]Goddard Space Flight Center, Greenbelt, MD, USA

[c]Space Telescope Science Institute, Baltimore, MD, USA

[d]Carnegie Institution of Washington, Department of Terrestrial Magnetism, Washington, DC, USA

**Abstract**

We consider the application of interferometry to measuring the sizes and shapes of small bodies in the solar system that cannot be spatially resolved by today's single-dish telescopes. Assuming ellipsoidal shapes, we provide a formalism to derive the shape parameters from visibility measurements along three different baseline orientations. Our results indicate that interferometers can measure the size of an object to better than 15% uncertainty if the limb-darkening is unknown. Assuming a Minnaert scattering model, one can theoretically derive the limb-darkening parameters from simultaneous measurements of visibilities at several different projected baseline lengths to improve the size and shape determination to an accuracy of a few percent. The best size measurement can be reached when one axis of the object's projected disk is aligned with one baseline orientation, and the measurement of cross-sectional area is independent of baseline orientation. We construct a 3-D shape model for the dwarf planet Haumea and use it to synthesize interferometric data sets. Using the Haumea model, we demonstrate that when photometric light curve, visibility light curve, and visibility phase center displacement are combined, the rotational period and sense of rotation can all be derived, and the rotational pole can be estimated. Because of its elongated shape and the dark red spot, the rotation of Haumea causes its optical photocenter to move in a loop on the sky. Our simulations show that this loop has an extend of about 80 µas without the dark red spot, and about 200 µas with it. Such movements are easily detectable by space-based astrometric interferometer designed e.g. for planet detection. As an example, we consider the possible contributions to the study of small bodies in the solar system by the Space Interferometry Mission. We show that such a mission could make substantial contributions in characterizing the fundamental physical properties of the brightest Kuiper Belt Objects and Centaurs as well as a large number of main belt asteroids. We compile a list of Kuiper Belt Objects and Centaurs that are potentially scientifically interesting and observable by such missions.

1.  Introduction

Directly measuring the sizes, rotations and densities of dwarf planets and other small bodies in the solar system has yielded fundamental insights into their interior structures and formation histories. For example, recent images of (1) Ceres from the Hubble Space Telescope (HST) showed that Ceres is an oblate spheroid with a polar radius of 454.7±1.6 km and an equatorial radius of 487.3±1.8 km, indicating an average density of 2,077±36 kg m$^{-3}$ (Thomas et al., 2005) with its mass measured elsewhere (Viateau and Rapport, 2001; Michalak, G., 2000). Note that often the uncertainty of size dominates the density determination. This new density determination suggests that Ceres has ~25% water by mass; together, the density and oblateness measurements indicate that Ceres has a differentiated interior. Based on the constraints provided by the observations, McCord and Sotin (2005) and Castillo-Rogez and McCord (2009) developed new models for the evolutionary history of Ceres. As one of the only few minor bodies in the asteroid belt that survived the violent collisions between solar system small bodies, Ceres is considered to be an embryo, analogous to the cores of giant planets. These new results have strong implications for the accretion and heating histories of the solar system's giant planets.

Outer solar system bodies are also ripe for new size and shape measurements. The Centaurs, the small bodies whose orbits cross those of one or more giant planets, make good targets for size and shape measurements. The shapes and densities of Kuiper Belt Objects (KBOs) have important implications for the physical conditions in the outer protoplanetary nebula. For example, Brown and Trujillo (2004) measured that the angular diameter of KBO (50000) Quaoar is 40.4±1.8 milliarcsecond (mas) using the ACS/HRC onboard HST. The accurate size measurements yielded a more accurate albedo and constrained the ability of this KBO to retain volatiles and sustain an atmosphere (Schaller and Brown, 2007). Brown et al. (2006) measured the angular diameter of (136199) Eris to be 34.3±1.4 mas using the same technique, confirming that Eris is larger than Pluto. These measurements can help determine which objects should be classified as dwarf planets.

Optical interferometry has emerged as a new technique for measuring the sizes and shapes of small bodies like asteroids, Centaurs, and KBOs. For instance, Delbo et al. (2009) measured the size of Asteroid (951) Gaspra with the Very Large Telescope Interferometer-Mid-infrared Interferometric Instrument (VLTI-MIDI). The VLTI-MIDI visibilities on this source, obtained from one baseline orientation at multiple infrared (IR) wavelengths, agree with the detailed shape model derived from Galileo flyby images (Thomas et al., 1994). Delbo et al. (2009) also obtained visibilities for Asteroid (234) Barbara at multiple wavelengths, and deduced that a binary model best fits the visibility data for this asteroid.

This paper discusses the application of optical interferometry to measuring the properties of dwarf planets, KBOs, and Centaurs, including their sizes, shapes, surface features, and rotations. We calculate generic models for the limb-darkening of these objects using triaxial ellipsoids and derive some fundamental principles for interferometric measurement of these bodies. We construct a 3-D shape model for the dwarf planet Haumea and use it to synthesize interferometric datasets. Finally we apply the results to the Space Interferometry Mission Lite (SIM) and discuss the capabilities of this proposed mission for planetary sciences.

## 2. Interferometry for solar system small bodies
*2.1 Standard measurement techniques*

Although some small bodies in the solar system, like Ceres and Quaoar, are resolvable, as we described above, most have maximum angular sizes below the resolution limits of the most powerful existing telescopes on the ground or in space. Instead, most existing size constraints come from combining radiometric measurements in the visible and in the thermal IR (see Harris and Lagerros, 2002, and references therein). The idea behind radiometry is that the brightness of an object in the visible wavelengths is determined by reflected sunlight, which is proportional to the cross-sectional area and the hemispherical albedo of the object. The absorbed solar flux heats up the surface of an object, which re-radiates in the IR. Therefore the IR brightness of the object is proportional to its cross-sectional area and absorption, which is unity minus the bolometric Bond albedo, $A$. Combining visible and IR measurements yields both parameters: albedo and size.

Although it appears simple on the surface, radiometry is complicated. For precise radiometry, one needs to measure the bolometric Bond albedo, $A$, i.e., the reflectance integrated over all wavelengths and directions, to determine the total energy that is absorbed and re-radiated in the thermal IR. Making this measurement is almost impossible; therefore, radiometry generally relies on models. Photometric models, often empirical, are usually invoked to predict the hemispherical reflectance at non-observable phase angles and wavelengths from the phase function and geometric albedo, $p_V$. These models often contain the simplifying assumptions that phase functions are linear in magnitude with respect to phase angle and independent of wavelength.

Radiometry also relies on thermal models like the refined standard thermal model (STM; Lebofsky et al., 1986), the near-Earth asteroid thermal model (NEATM; Harris, 1998), and the fast rotating thermal model (FRM; Harris and Lagerros, 2002; Delbo and Harris, 2002, and references therein). Thermal models are also needed to describe the effect of macroscopic roughness and thermal inertia. These thermal models relate the measured total thermal IR flux (usually just at one observing geometry for part of the surface) to surface temperature distribution, and relate the equilibrium surface temperature distribution to albedo and size. The thermal balance on the surface of a planetary body depends on its shape and rotational state, both usually unknown. Because of the many uncertainties in the models, radiometric size measurements typically have uncertainties of 20% or higher.

If an object is large and at least marginally resolved, then its size can be directly measured from imaging data, though this method is not entirely model- or assumption-free. For this purpose, one has to understand the characteristics of the imaging instrument and accurately determine the point-spread-function (PSF) and the sub-pixel location of the object to accurately model the images of the object. More importantly, understanding the images requires modeling the object's surface brightness distribution including limb-darkening, which is generally unknown and often translates into 15% or higher uncertainties in the inferred sizes.

Rotational lightcurve measurements of how objects vary in total brightness as they rotate have been the most frequently used method to constrain the shapes and rotational states of spatially unresolved small bodies. The amplitudes of rotational lightcurves yield the lower limits on the axial ratios of the objects, while the periods of lightcurves correspond to the rotational periods of the objects (e.g., Sheppard et al., 2008).

There are several limitations to this method. First, although the shape of an object usually dominates lightcurves (e.g., Kaasalainen et al. 2002), for objects that are nearly rotationally symmetric, albedo variations on the objects' surfaces can have a significant effect. Pluto (Buie et al., 1997), Ceres (Li et al., 2006), and Asteroid (4) Vesta (Binzel et al., 1997; Li et al., 2010) all have substantial albedo variations on their surfaces. Lacerda et al. (2008) interpreted the unequal peaks and valleys in the rotational light curve of KBO (136108) Haumea as a dark spot on a triaxial shape.

Second, viewing angle affects the amplitudes of rotational lightcurves. Viewed pole-on, an object will show a flat rotational lightcurve no matter what shape it has unless there are strong shadows. When the amplitude of a rotational lightcurve is not large enough, the rotational period cannot be reliably measured. One generally needs rotational lightcurves from at least two significantly different viewing angles with respect to the pole orientation to unambiguously determine the axial ratio and rotational pole of an object. Lightcurve inversion techniques used to deduce the shapes of asteroids (Kaasalainen et al., 2002) assimilate observations from as many different observing and illumination geometries as possible. However, these techniques generally fail for KBOs, which are >30 AU from the Earth and have orbital periods measured in centuries, thus the viewing geometry varies very slowly.

With their faintness, small size, and small range of phase angles, KBOs are particularly problematic for size and shape measurements with single-dish telescopes. Another technique to measure the sizes of asteroids and satellites of planets is occultation (e.g., Tanga and Delbo, 2007, and references therein). But for slow moving KBOs and their orbits poorly determined from the limited length of observed arc, the chance of observing occultation is not good. With currently available observing capabilities, the measurements of the sizes and albedos of most KBOs are only good to at best 20%.

*2.2 Interferometry in planetary sciences*

Compared with traditional observing facilities, interferometers provide us with extraordinarily high spatial resolution, and enable direct determination of the sizes, shapes, and rotations of small bodies in the solar system with better accuracy. Sizes derived from interferometry can complement and refine radiometrically-determined sizes because they depend on different model parameters.

The Fine Guidance Sensor (FGS) onboard HST is a space-based optical interferometer. Hestroffer et al. (2002) and Tanga et al. (2003) used the HST/FGS to measure the fringes of several asteroids to search for binaries. Their attempt suggested that space-based interferometry could provide direct measurements of the shape, size, and rotation of asteroids. However, HST/FGS is never a dedicated science interferometer. In our work, we discuss a general case of applying interferometry to solar system extended sources.

The resolving power of an interferometer is characterized by its fringe spacing, $\lambda/B_\perp$, where $\lambda$ is the wavelength, and $B_\perp$ is the length of baseline projected onto the sky plane. The Atacama Large Millimeter/sub-millimeter Array (ALMA), the largest interferometric array currently under construction, has an angular resolution of about 5 mas in its most extended configuration. On the other hand, ALMA will operate in sub-millimeter wavelengths where small bodies are faint. Lovell (2008) discussed the general capability of ALMA in observing asteroids; Busch (2009) considered using the high spatial resolution of ALMA to map the shapes and large-scale surface features of main belt asteroids. Both studies show that ALMA is able to spatially resolve and map the thermal emission from the surfaces of the largest asteroids to study their thermal properties.

Of particular interest for our study is the proposed optical-wavelength Space Interferometry Mission (SIM-Lite; Unwin et al., 2008; Davidson et al., 2009). SIM has two interferometers with the same baseline orientation and lines-of-sight (LOS). The "science" interferometer has a 6- meter baseline, and the "guide" interferometer has a 4.2-meter baseline. Both interferometers can measure the interferometric fringes at 80 narrow-band channels simultaneously from 450 nm to 900 nm, with the capability of binning several channels on the detector in order to increase the signal-to-noise ratio if needed. Therefore, for any observations the observer will automatically obtain the total flux, visibility amplitude, and phase center of the target relative to reference stars over the whole spectral range of SIM. The fringe period of SIM varies from a few to about 50 mas. The system is specially designed for high-precision astrometry, and is hence capable of measuring fringe phase to micro-arcsecond (µas) precision. SIM can be rotated around the line of sight in order to measure the complex fringe visibility along various baseline orientations; the typical slew time to rotate the baseline by 90º is of order tens of minutes. Thanks to its location in space above the disturbing atmosphere of the Earth, SIM can measure fringe amplitudes to ~1% and fringe phases to one part in $10^3$ on distant stars in an hour of integration at a limiting magnitude of *V*~20 mag. The sensitivity can be further improved with additional integration; the limiting astrometric precision (set by instrument instability) is better than 1 μas.

Another future interferometer that may be useful for this kind of research is the non-redundant mask interferometer on the James Webb Space Telescope (Sivaramakrishnan et al., 2009), part of the Fine Guidance Sensor Tunable Filter Imager (FGS-TFI). In general, the spatial resolution of interferometers is about one order of magnitude smaller than the minimum size directly measurable by current single-dish telescopes. For comparison, the highest-resolution instrument currently operating on HST, the Wide Field Camera 3 (WFC3), can only resolve objects >80 mas (2 pixels). Ground based observations for faint objects like KBOs are usually limited by atmospheric seeing, which is at best several hundred mas.

We will use the example of SIM as a prototype of space-based optical interferometer whenever the need arises in this paper. The extremely high fringe phase stability is only achievable in space. In principle, the basic conclusions about using interferometric visibility to determine the fundamental properties of solar system objects apply to any wavelength. The determining factor will be the diameters of the targets measured in terms of $\lambda/B_\perp$. With a 6-meter primary mirror and IR bandpass, JWST will have interferometric resolution a few times lower than that of SIM. ALMA, on the other hand, works in sub-millimeter to millimeter wavelengths but with much longer baselines (0.15 – 18 km) (Tarenghi, 2008), providing a $\lambda/B_\perp$ resolution as small as 5 mas.

*2.3 KBOs and Centaurs*

To understand the capability of interferometers like SIM in observing solar system small bodies, we examined 109 KBOs and 26 Centaurs that have published absolute magnitudes and/or size measurement from infrared (IR) radiometry (Romanishin and Tegler, 2007; Sheppard et al., 2008; Stansberry et al., 2008). For the objects without size estimates, we assumed a visible geometric albedo, $p_V$=0.1 for objects smaller than 600 km in diameter, and $p_V$=0.6 for objects larger than 600 km, and estimated their sizes from their absolute magnitudes. Fig. 1 shows the sizes and heliocentric distances of all 135 objects we considered.

In Fig. 1, the yellow and green regions mark the solar system objects that can be measured by the 6-m science interferometer of SIM, assuming a limiting *V*-magnitude of 20. For distant objects such as KBOs, the apparent brightness determines whether or not they are observable with SIM; for nearby objects, angular size is the limiting factor (Section 3). The transition occurs between 5 and 20 AU, depending on the object's albedo. The largest KBOs like Eris (Brown et al., 2006), (136472) MakeMake (Stansberry et al., 2008), and Haumea (Lacerda and Jewitt, 2007) start to be over-resolved by SIM. In this case, the 4.2-m baseline guide interferometer on SIM might also be harnessed for science observations (Section 7.2). This additional baseline would provide a fringe spacing larger by almost 50%, making more objects accessible to SIM as marked by the green and blue regions in Fig. 1.

The region occupied by objects that are accessible to ALMA is also marked in Fig. 1 by the dotted area for a comparison with SIM. For our calculation, we assumed that an object is in instantaneous thermal equilibrium with solar radiation, has a uniform surface temperature, and we used a continuum sensitivity of 0.7 mJy at 675 GHz for ALMA (Johnstone et al., 2010).

Fig. 1 shows that of the 135 KBOs and Centaurs that we considered, SIM can observe 55 KBOs and 26 Centaurs, while ALMA can observe somewhat fewer because of its sensitivity limits: 42 KBOs and 18 Centaurs. SIM generally performs better than ALMA for objects at 5 AU to 20 AU, where Centaurs reside.

Table 1 lists 38 objects in the outer solar system that would be good candidates for observation with SIM, including 21 KBOs and 17 Centaurs. These objects are plotted as stars in Fig. 1. The largest KBOs, such as Eris, Pluto, and MakeMake are difficult for SIM to observe even with the 4.2-m guide interferometer because of their large angular sizes, and are not listed. Of the listed objects, about half are larger than the fringe spacing at 0.4 μm with SIM science interferometer. All of them can be observed with the help of the guide interferometer. At the other extreme, some of the smallest Centaurs are too small to observe at the red end of SIM spectral range, but easily resolvable in the blue end. Stansberry et al. (2008) reported radiometric sizes for most of the objects listed. The radiometry incorporates only one or two IR bands, and the radiometric size uncertainties are typically >20%. SIM is expected to achieve much better precision (Section 3).

About half of the objects have photometric lightcurves with photometric ranges smaller than 0.1 mag, making it hard to measure their rotational periods from their photometric lightcurves alone. Even for those objects with photometric ranges large enough to allow us to fit rotational periods, it is hard to tell whether the lightcurves are single- or double-peaked (Table 1) except for some special cases like Haumea (Lacerda et al., 2008). For photometric lightcurves with ranges less than 0.2 magnitudes, albedo markings and non-spherical shapes usually both contribute to the shape of lightcurves (Sheppard et al., 2008). Interferometric measurements would allow us to distinguish the effects of elongation from albedo markings in photometric lightcurves and to determine their rotational states (Section 4). Visibility phase measurements are extremely sensitive to albedo features on the surface even if those features only causes a few percent change in total brightness (Section 6).

*2.4 Asteroids*

The first VLTI-MIDI measurements of the sizes of Gaspra and Barbara demonstrated the feasibility of applying interferometric measurements to asteroids (Delbo et al., 2009). SIM has much higher sensitivity than VLTI-MIDI, and should be able to observe much fainter (smaller) asteroids not only in the main asteroid belt, but also in other populations such as near Earth asteroids, Jovian Trojans, etc. Fig.1 shows that at 2-3.5 AU, the distances of main belt asteroids (MBAs), SIM can be used to measure objects down to several km in size.

Measuring of size of MBAs can help us understand the collisional process that played a major role in the evolution of asteroids. Besides the overall MBA size distribution, it would be informative to measure the size distributions in the dynamical families of asteroids, groups of asteroids formed together in recent collisions (Davis et al., 2002). The rotational states of MBAs contain substantial information about their origin, internal structure, and evolution (Pravec et al., 2002). Understanding the rotational states of asteroids in dynamical families could constrain their collisional histories and the properties of the parent bodies.

## 3. Simple analytic visibility models
*3.1 Circular disk model*

Visibility is the Fourier transform of the brightness distribution of the source divided by the total intensity. Consider the simplest case of a circular disk with a uniform brightness distribution and an angular diameter of $\theta$. The normalized visibility, $V$, of such a disk measured by an interferometer with a projected baseline, $B_\perp$, at wavelength, $\lambda$, is given by a jinc function,

$$V = \frac{2J_1(\pi\theta B_\perp/\lambda)}{\pi\theta B_\perp/\lambda}, \qquad (1)$$

where $J_1$ is the first-order Bessel function of the first kind. The normalized visibility is defined to be unity at $\theta=0$.

If the diameter of the circular disk is smaller than where the first zero of jinc function occurs, $\theta = 1.22\ \lambda/B_\perp$, then the size can be uniquely determined by inverting Eq. (1) from a single visibility measurement. If the size of the disk exceeds $1.22\ \lambda/B_\perp$, then the inverse yields multiple solutions. In this case, one would need other constraints, such as those from radiometric measurements, to uniquely determine the size of the disk.

The uncertainty of the size measurement is determined by both the measurement uncertainty of the visibility and slope of the jinc function at the measured visibility value. The sweet spots are where the diameters fall in the steepest slopes of the visibility curve, roughly between 0.2 and 1.2 $\lambda/B_\perp$ if only the primary beam is considered. In this regime, the relative uncertainty of the measured size will be comparable to the uncertainty of visibility measurements. SIM has an expected accuracy of 1% for the correlated flux, and the corresponding uncertainty of visibility can be up to 10% for a visibility higher than 0.1. When the size is larger than the primary beam of the jinc function, the relative uncertainty of the measured size will be higher due to the flatter nature of the curve.

In this paper, we will restrict ourselves to objects with diameters less than $1.22\ \lambda/B_\perp$. In principle, one should be able to generalize our discussions to larger objects if other observations can provide additional constraints so that a unique solution can be determined. The uncertainties in the measurements of larger objects will be higher because of the mismatch between the fringe size and object size.

*3.2 Elliptical disk model*

Solar system small bodies are rarely spherical. Among 40 KBOs studied by Sheppard et al. (2008), about 29% have light curve amplitude greater than 0.15 mag, and about 12% have amplitudes greater than 0.60 mag. Examples of likely non-spherical objects include KBOs (20000) Varuna, which has a lightcurve amplitude of 0.42±0.02 mag and probably is rotationally elongated (Jewitt and Sheppard, 2002), 2001 QG$_{298}$, which has a lightcurve amplitude of 1.14±0.04 mag and is probably a contact binary (Sheppard and Jewitt, 2004; Takahashi and Ip, 2004), and Haumea, which is also rotationally elongated.

To take the next step beyond the circular disk model described above, we considered a triaxial ellipsoidal shape model. A triaxial ellipsoid with a dark albedo and color marking yields a good fit to the rotational light curves of Haumea (Lacerda et al., 2008). This kind of model can help to relate interferometric data with limited (*u*, *v*) coverage to the sizes, shapes, and rotational properties of solar system bodies without burdening the observer with too many potentially degenerate parameters.

For a triaxial ellipsoid, the projected shape is an elliptical disk. To uniquely determine the shape of the elliptical disk in the sky plane, one needs to measure three parameters: the semi-major axis, *a*, the semi-minor axis, *b*, and an angle $\varphi$ measuring the direction of the long-axis with respect to a reference direction, e.g., a baseline orientation in sky plane. The normalized visibility of such a uniform elliptical disk is

$$V = \frac{J_1(2\pi Q)}{\pi Q} \quad (2)$$

where $Q^2 = (a\cos\varphi)^2 + (b\sin\varphi)^2$. Here *a* and *b* are expressed in units of $\lambda/B_\perp$, which thereby does not appear in the equation. *Q* can be considered the effective angular radius of the projected ellipse along the orientation of the corresponding baseline (Hestroffer et al., 2002). Eq. 2 shows that with this model, each visibility measurement can be interpreted as yielding a measurement of *Q*.

We find that three visibility measurements along three non-degenerate baseline orientations suffice to uniquely determine the uniform ellipse model. To illustrate how this works, let us consider two special cases. In the first case, the three effective radii $Q_i$ (*i*=1,2,3) are measured from three baselines separated by 45° (Fig. 2a, Eq. 3). Let us assume that $\varphi$ is the angle between the long axis of the elliptical disk and the first baseline orientation (corresponding to $Q_1$). For this geometry, the three elliptical disk parameters ($a$, $b$, $\varphi$) can be written in terms of $Q_i$ as

$$\begin{cases} (Q_1^2 - Q_2^2)\sin(2\varphi) = (Q_1^2 + Q_2^2 - Q_3^2)\cos(2\varphi) \\ a^2 = \frac{1}{2}\left[Q_1^2 + Q_2^2 + \frac{Q_1^2 - Q_2^2}{\cos(2\varphi)}\right] = \frac{1}{2}\left[Q_1^2 + Q_2^2 + \frac{Q_1^2 + Q_2^2 - 2Q_3^2)}{\sin(2\varphi)}\right] \\ b^2 = \frac{1}{2}\left[Q_1^2 + Q_2^2 - \frac{Q_1^2 - Q_2^2}{\cos(2\varphi)}\right] = \frac{1}{2}\left[Q_1^2 + Q_2^2 - \frac{Q_1^2 + Q_2^2 - 2Q_3^2)}{\sin(2\varphi)}\right] \end{cases} \quad (3)$$

In the second case, the three baselines are 60º apart from one another (Fig. 2b). In this case, we find

$$\begin{cases} (2Q_1^2 - Q_2^2 - Q_3^2)\sin(2\varphi) = \sqrt{3}(Q_3^2 - Q_2^2)\cos(2\varphi) \\ a^2 = \frac{1}{3}\left[Q_1^2 + Q_2^2 + Q_3^2 + \frac{2Q_1^2 - Q_2^2 - Q_3^2}{\cos(2\varphi)}\right] = \frac{1}{3}\left[Q_1^2 + Q_2^2 + Q_3^2 + \frac{\sqrt{3}(Q_3^2 - Q_2^2)}{\sin(2\varphi)}\right] \\ b^2 = \frac{1}{3}\left[Q_1^2 + Q_2^2 + Q_3^2 - \frac{2Q_1^2 - Q_2^2 - Q_3^2}{\cos(2\varphi)}\right] = \frac{1}{3}\left[Q_1^2 + Q_2^2 + Q_3^2 - \frac{\sqrt{3}(Q_3^2 - Q_2^2)}{\sin(2\varphi)}\right] \end{cases} \quad (4)$$

The propagation of the measurement uncertainties in the three visibilities into the uncertainties of the size parameters, $a$ and $b$, and the orientation parameter, $\varphi$, can be characterized by the derivatives of above expressions with respect to visibilities. We studied the uncertainties of the three parameters, $a$, $b$, and $\varphi$, assuming a visibility uncertainty of 2% (if measured by SIM for visibilities higher than 0.5). Skipping some tedious algebra, we summarize our findings as following:

a) The uncertainty of size measurement depends on the orientation of the baselines. The best accuracy is achieved with the first configuration (45º baseline separation) in the special case when the axes of an elliptical disk align with the two perpendicular baselines.
b) If both $a$ and $b$ of the disk are within 0.2 to 1.2 $\lambda/B_\perp$, then they can be measured to roughly 2% uncertainty in the best case, similar to uncertainty of the size of circular disk.
c) The accuracy of the measured orientation, $\varphi$, strongly depends on the axial ratio of the disk. When the lengths of two axes differ by 10% or more, $\varphi$ can be measured to better than 5º.
d) The uncertainty of $\varphi$ only weakly depends on the orientations of baseline.

e) The cross-sectional area of the disk, *πab*, can always be measured to roughly 2%, independent of the orientations of baselines.

In an actual measurement, one might wish to optimize the choice of baseline orientations to maximize the likelihood of obtaining the best measurement with the smallest uncertainty and the minimum observing time requirement based on any *a priori* knowledge of the object according to above rules. For example, if the orientation of the axes of the disk has been constrained, then one may want to choose the first baseline configuration discussed above, and align the two perpendicular baselines with the axes so that the size of the object can be measured in two measurements with minimal uncertainties.

*3.3 Deducing the 3-D shape*

Three visibility measurements along different baseline orientations at the same rotational phase determine the shape and size of the projected elliptical disk of an ellipsoidal object. As the object rotates, its projected shape from various rotational angles can be measured. In principle these measurements allow the observer to deduce the object's 3-dimensional shape (Hestroffer et al., 2002). The reconstruction of 3-D triaxial ellipsoidal shape from projection is always solvable from forward simulations (Gendzwill and Stauffer, 1981). For more complicated shapes, one should in principle be able to construct the simple convex-hull (no concavity is included) 3-D shape from the 2-D projections in three rotational phases, or two if the pole orientation is assumed to align with the short-axis of the object.

**4. Rotational states from visibility**

To further investigate the measurements of KBO and Asteroid properties with interferometry, we employed an asteroid-shape-to-image (ASTI) tool to simulate the images of an asteroid from its 3-D shape model. The ASTI tool takes any 3-D shape model composed of triangular facets, calculates the reflectance of each facet according to photometric properties and illumination and viewing angles, and generates synthetic images. Shadows and variations of photometric properties can be included in the model. The ASTI tool can also be used to calculate thermal models of an asteroid and to synthesize its thermal images (thermal-ASTI, or TASTI tool). It has been successfully used to analyze the disk-resolved photometry of several asteroids and cometary nuclei (Li et al., 2004; 2006; 2007a; b; 2009; 2010). We used the ASTI tool to generate the synthetic data sets for the dwarf planet Haumea in this work.

We note that visibility could be sensitive to the actual (irregular) shape and brightness distortion caused by concavities. However, for solar system objects of several hundred km or larger in diameter, their shapes are in general close to triaxial ellipsoid (Tanga et al., 2009). Large KBOs should contain minimal concavities. In this work we discuss the general principles of using interferometry to solar system objects with emphasis on relatively large Centaurs and KBOs, and focus our simulations on triaxial ellipsoidal shapes.

*4.1 Visibility lightcurves*

When an object rotates, depending on its shape and the angle between pole orientation and the LOS, as well as its surface albedo variations, its total brightness may vary, generating a lightcurve. Most often, when we know the rotational period or pole orientation of an asteroid, it has been deduced from lightcurves (Kaasalainen et al., 2002). Lightcurves for KBOs are also useful, though photometric lightcurves with less than 0.1 mag photometric range are not uncommon for faint KBOs that have photometric uncertainties of about the same level (e.g., Sheppard et al., 2008).

The rotation of an object may also cause periodic change of its visibility; this time variation can contain substantial information that is not carried by ordinary lightcurves. In our discussion, we will call the traditional brightness lightcurves "*photometric lightcurves*". By analogy, we will call the rotational variation of an object's visibility amplitude along a particular baseline a "*visibility amplitude lightcurve*", or simply "*visibility lightcurve*".

Visibility lightcurves contain spatial information along the baseline orientations that they are measured on, so can overcome many limitations of photometric lightcurves. For example, it is usually impossible to distinguish between a rotationally symmetric (oblate) object and an elongated object viewed nearly pole-on from photometric lightcurves. A visibility lightcurve has the potential to break this degeneracy.

A precaution has to be made that, in all of our analysis, we assumed no rotational variation on the spectrum of an object, but albedo variations. Variation in spectrum usually indicates changes in the distribution of brightness on the disk, thus causing variations in visibility. Almost all solar system objects show spectral variations along rotation to some level. But for many objects, the spectral change is at a few percent level and can be ignored. SIM provides spectral information simultaneously with visibility information (section 7.1), so with SIM one will be able to justify or reject this assumption for each case.

*4.2 Haumea models*

To show how visibility lightcurves can reveal an object's rotational states, we used the ASTI tool to simulate the photometric lightcurves, the visibility amplitude lightcurves, and the visibility phase for SIM measurements of the dwarf planet Haumea. Rabinowitz et al. (2006) and Lacerda et al. (2008) determined the properties of this KBO from its unusually fast rotation. Its high albedo of $p_V > 0.6$ is consistent with a surface composed of almost pure crystalline water ice (Trujillo et al., 2007). Strong, but not extreme, limb-darkening (see next section for the discussions on limb-darkening) is expected on its surface. We therefore assumed a Minnaert *k* parameter of 0.85 to model its limb-darkening (see below). We adopted the axial ratios of Haumea: $b/a = 0.86$ and $c/a = 0.54$, where *a*, *b*, and *c* are the semi-major, semi-intermediate, and semi-minor axes (Lacerda and Jewitt, 2007), respectively. We assumed that the rotation of Haumea has fully relaxed and its short-axis is aligned with its rotational axis. For every 5º of rotation, we simulated the disk-resolved image by calculating the brightnesses of all pixels on the disk according to their scattering geometries. To generate the photometric lightcurves, we integrated the brightness in the images. To calculate the visibility lightcurves, we evaluated the Fourier Transforms of the simulated images at the appropriate baseline and normalized by the total flux.

Fig. 3 shows our model for Haumea at two viewing angles. The first viewing angle has the LOS oriented 80º from the short axis (almost edge-on, upper panel). This view is probably close to the actual viewing geometry of Haumea (Lacerda and Jewitt, 2007). The second viewing angle has the LOS oriented 15º from the short/rotational axis (almost pole-on, lower panel). We calculated the photometric lightcurves and visibility of this object along the two baselines shown in Fig. 3, one 20º from the projected orientation of pole (B1), another perpendicular to the first one (B2). At first, we assumed a uniform surface; we discuss albedo variations in Section 6. The simulated photometric lightcurves, visibility amplitude lightcurves, and visibility phase center displacement for both viewing angles are shown in Figs. 4, 5, and 6.

From the simulations, we will show that the rotational state, including period, pole orientation, and the sense of rotation, can all be determined from the combination of photometric lightcurve, two visibility amplitude lightcurves, and the track of visibility phase center.

*4.3 Rotational period*

Let us first consider the near edge-on viewing angle as shown in Fig. 4 by the thick, solid (B1) and dashed (B2) lines in the left panels and in Fig. 5a. For this case, the object shows a double-peaked photometric lightcurve with peaks of equal height and a photometric range of about 0.23 mag. The peaks and the valleys correspond to the maximum and minimum cross-sectional areas, respectively. The visibility lightcurves are also double-peaked with equal heights, with an average of 0.68 and a range of 0.03 (~4.4% of average) along B1, and an average value of 0.50 and a range of 0.06 (~12%) along B2. The maxima of visibility amplitude lightcurves occur near, but not exactly at, the photometric lightcurve maxima and minima.

Fig. 5a shows the trace of visibility phase center in the sky for the first viewing geometry (nearly edge-on). Visibility phase center is measured relative to the mean of nearby reference stars to an accuracy of ~10 μas. SIM can access a 7º star-field without moving its baseline. This guarantees that even for relatively fast moving main belt asteroids, SIM can still precisely measure the phase center over at least one rotation. For objects with rotations relaxed to their principle axes, we expect the trace of visibility phase center relative to geometric center to close a loop in the course of a rotation. Therefore, the proper motion of an object can be separated from the rotation-induced phase center shift by fitting a smooth line to the overall trace of phase center relative to mean star background both in space and in time. The relative offset of visibility phase center to the geometric center of the target, as shown in Figs. 5 and 6 can then be recovered. The precisely determined proper motion of the target, as a byproduct, can be used to refine the orbit of the target.

The plots in Fig. 5 are tilted such that the two axes align with the two baselines. During a rotation, the phase center closes two loops in the path counterclockwise. The loop is about 80×40 μas in angular size, elongated in roughly the east-west direction, which is the direction of the long-axis of the disk. The thick line in Fig. 6 shows the case of nearly pole-on viewing angle, where the size of the loop is much smaller, only about 20×10 μas. Figs. 5 and 6 dramatically illustrates the extremely high astrometric precision provided by SIM.

It is relatively straightforward to measure the rotational period from visibility lightcurves. The techniques for inferring a rotational period from photometric lightcurves, like the phase dispersion technique (Stellingwerf, 1978), can be readily adapted to visibility lightcurves. If photometric lightcurves covering a long time span are available, then the rotational period can be improved by fitting all photometric lightcurves at once. E.g., Chamberlain et al. (2007) improved the precision of the rotational period of Ceres by almost three decimal points to 7 milliseconds by combining the photometric light curves of Ceres that spanned roughly 50 years. Similar procedures can be applied to visibility lightcurves measured over a long time baseline. In this process, one not only needs to take into account the possible changes in viewing angles, but also the baseline orientations projected into sky. Note, however, that if the determination of rotational period is the only purpose, then it is much easier to obtain photometric lightcurves given that the current observing techniques can often reach 1% precision in relative photometry.

*4.4 Pole orientation*

Visibility lightcurves also enable the determination of pole orientation from observations at only a single viewing geometry. The range of photometric lightcurve is related to the change of cross-sectional area, and the range of visibility amplitude lightcurve is related to the change of size along the interferometer baseline, both obeying the same limb-darkening. Therefore, by comparing the ranges of photometric lightcurves and visibility amplitude lightcurves, one will be able to determine the projected orientation of the rotational pole in sky plane.

Let us take the simulations of Haumea as an example, as shown in Figs. 3a, 4, and 5a, and try to interpret the lightcurves using a triaxial spheroid model. For the near edge-on view shown in the left panels of Fig. 4, the range of photometric light curve reflects a ~24% change in the effective projected cross-sectional area. The range of visibility amplitude lightcurve along B1 is 0.03, reflecting a ~5.5% change of the effective size along that baseline according to Eq. (2). The range of visibility amplitude along B2 is 0.06, reflecting a ~7.4% change of effective size along that baseline. With an expected 1% uncertainty of correlated flux, SIM can distinguish a visibility change of 0.01, enough to observe the small amplitudes of visibility lightcurves we discuss here.

If we assume a uniform surface albedo and scattering phase function, the range in the photometric lightcurve indicates that $a/b>1.24$. If one of the baselines were aligned with the projected direction of the pole, the range of visibility amplitude light curve would be $\sim a/b$ times a factor determined by $c$ and the angle between the LOS and the rotational axis. The facts that both ranges are non-zero and smaller than the $a/b$ ratio suggested by photometric lightcurve imply that neither baseline is aligned to the projected direction of pole. Moreover, the visibilities show that the maximum size along B2 is longer than that along B1 by 31%. Therefore, we can infer that the projected direction of pole must be closer to B1.

On the other hand, if Haumea is viewed nearly pole-on (right panels of Fig. 4), then the photometric range of its lightcurve is only about 0.05 mag, but the change of the size along each of the two baselines is ~7% as derived from the visibility range of 10% with an average of 0.55. The visibility lightcurves in this case reveal the true axial ratio of the object. The nearly opposite phasing of the two visibility amplitude light curves suggests that its rotation is almost pole-on.

*4.5 Sense of rotation*

Depending on observing geometry, limb-darkening (see next section) on an elongated body and albedo variations may both potentially cause displacement of an object's photocenter from its geometric center. Our simulated cases suggested that the photocenter displacement introduced by limb-darkening can be up to 20% of the radius of the disk. The highlight (not the dark area) in Fig. 3 illustrates photocenter displacement for an extremely strong limb-darkening. The displacement of the photocenter causes a shift in the phases of complex visibilities from the geometric center, and a variation in the phase center as the object rotates.

Our model of Haumea shows that the motion of the phase center indicates the sense of rotation (prograde vs. retrograde), and helps constrain the pole orientation. Since the brightest spot on an object is always near the sub-solar and sub-Earth point if the phase angle is small, which is true for all KBOs, visibility phase indicates the projected path of the sub-solar point relative to the geometric center. Figs. 5 and 6 show the motion of the phase center as a function of time for a nearly edge-on view and a nearly pole-on view, respectively. The phase center describes a closed loop on the sky. The orientations of the two elliptical paths do not align with the actual path of the sub-solar point, which would be horizontal or vertical in the plot. This discrepancy arises from limb-darkening and/or the non-alignment of baselines with the axes of the object. The absolute value of visibility phase does not change the shape of the tracks of photocenter, therefore does not affect the discussions in this section.

As indicated by the photometric lightcurve, for the edge-on view at rotational phase 0.0, the long-axis of the object rotates away from the LOS direction. Since the sub-solar point follows the long-axis at this phase, the direction of motion of the sub-solar point implies the direction of the rotation of long-axis, and therefore the sense of rotation. The path of the photocenter puts constraints on the pole orientation, too. The nearly edge-on view results in an elongated shape for the photocenter path, while the nearly pole-on view results in a small, almost circular loop.

In the analysis discussed in this section, we include limb-darkening but we did not consider any albedo features, which could result in completely different lightcurves and visibility phase center displacement, as shown by the thin (both solid and dashed) lines in Figs. 4 and 6 as well as in Fig. 5b. We will show in Section 5 that different limb-darkening parameters would not affect these conclusions. We will discuss in Section 6 how albedo features can be studied from various lightcurves.

**5. Limb-darkening**

The sizes of stars measured from interferometry must be corrected to account for stellar limb darkening (e.g., Quirrenbach et al., 1996; Hajian et al., 1998). One must correct interferometric measurements of the size of planetary objects for limb darkening effects as well. A limb-darkened body has a higher visibility than a uniform disk of the same physical size, causing the best-fit uniform disk model to underestimate the size.

*5.1 Limb-darkening models*

Limb darkening of planetary bodies likely originates from surface roughness, from small scales features that are just big enough to cast shadows, to large scale topographical features that are just below the resolution limit of images (e.g., Hapke, 1993; Shepard et al., 1998). Bright objects with icy surfaces, such as those of Europa, Enceladus, Triton, etc., usually have strong limb darkening, completely determined by the local solar elevation. For dark surfaces like the Moon's, the limb-darkening is generally weak, and can be absent at opposition.

Empirical models have been developed to describe the limb-darkening profile of planetary objects. For the models in this paper, we adopted the Minnaert (1941) limb-darkening model. In this model, the bi-directional reflectance $r$ is given by $r \propto \mu_0^k \mu^{k-1}$, where $\mu_0 = \cos(i)$ is the cosine of incidence angle, and $\mu = \cos(e)$ is the cosine of emission angle. A single parameter, $k$, determines the degree of limb-darkening of a surface; $k = 1$ represents the limb-darkening of the extreme case of a bright surface with 100% reflectance, while $k = 0.5$ represents a uniform disk when viewed at opposition (0º solar phase angle), typical for dark surfaces.

Another widely adopted semi-empirical photometric model is the Hapke model (Hapke, 1993, 2002), which can yield some physical interpretations for its photometric parameters under reasonable physical assumptions. This model interprets the light scattering properties of a planetary surface in terms of fundamental physical properties, such as single-scattering albedo (SSA hereafter). Some ingredients in this model, such as the single-particle phase function, employ empirical parameters relying on laboratory measurements of planetary surface analogues. The physical assumptions of the Hapke model, as well as the physical interpretations of the photometric parameters, have been the subjects of debates for decades (e.g., Shepard and Helfenstein, 2007). Compared with empirical models, the Hapke model always involves more parameters, and often a unique solution to the model is hard to find from limited, disk-integrated observational data. However, McEwen (1991) showed that, when the phase angle is less than about 30º, any surface that can be described by a set of Hapke parameters can be also described just as well by a unique Minnaert parameters, $k$. Therefore Minnaert model is sufficient to describe the limb-darkening of KBOs and MBAs observed from the ground or in the vicinity of the Earth.

*5.2 Measuring limb-darkening from visibility*

Fig. 7 shows the visibility of a spherical atmosphereless object viewed at opposition, for various degrees of limb-darkening, characterized by the Minnaert $k$ parameters ranging from 0.5 (uniform disk) to 1.0 (perfectly scattering surface). For a disk with a uniform brightness distribution ($k$=0.5) observed at low phase angle, the visibility is a jinc function as shown by the solid line in Fig. 7. The visibility of a limb-darkened disk ($k$>0.5) has a different dependence on the size of object. For an object with diameter near $\lambda/B_\perp$, the visibility can differ by almost a factor of 2 depending on the limb-darkening.

If the visibilities of an object can be obtained simultaneously over a range of $\lambda/B_\perp$, then we can use Fig. 7 to estimate the limb-darkening parameter of an object. Suppose visibilities at $(\lambda/B_\perp)_i$, where $i$=1, 2, …, $n$, are simultaneously measured to be $V_i$. By assuming a uniform disk model ($k$=0.5), one can calculate an initial guess, $x$, for the angular diameter of the object from one visibility, then plot the measured visibilities $V_i$ on Fig. 7 as a function of $x/(\lambda/B_\perp)_i$. The visibilities, $V_i$, will not fit the solid line in Fig. 7 unless the disk is indeed uniform. The task now is to find a new size estimate such the visibility curve fits the measured $V_i$'s. To find this new size estimate, simply slide the measured $V_i$ data along the $x$-axis in Fig. 7, which has a logarithmic scale, to find a curve that best fits the data. This method will simultaneously yield the limb-darkening parameter, $k$, and the correct size. It applies to data sets spanning both a range in wavelength and a range in baseline length – as long as you assume that the limb-darkening parameter is constant over the range of wavelengths used for the first case.

At thermal IR wavelengths, solar-system bodies likely have different limb darkening parameters caused by different physics than what applies at visible wavelengths. The idea of combining multi-wavelength data to deduce the limb-darkening parameters may not work across visible and thermal IR wavelengths. However, visibility measurements at thermal and visible wavelengths can be compared to study how the limb darkening varies with wavelength with the help of appropriate thermal models. Radiometric models often invoke a thermal beaming parameter, which characterizes the temperature distribution and the possible anisotropy of the thermal radiation, determined by surface roughness, thermal inertia, and rotational period (e.g. Harris, 1998; and reference therein). Visibility observations of solar system bodies in the IR may provide an independent assessment of the roughness and the beaming parameter, and help calibrate radiometric methods.

*5.3 Size measurement for limb-darkened objects*

If the visibility data have too little range of $\lambda/B_\perp$ or too much noise for determining the limb-darkening with the above method, then we can use our *a priori* knowledge of the object's limb-darkening to correct for the measured size, assuming a uniform disk. Here we will discuss how the correction factor, *f*, would depend on the Minnaert *k* parameter.

Let us start by defining the *visibility size* as the size of an object measured from a visibility via Eq. 1, which assumes a uniform, circular disk. The actual size is the visibility size times the correction factor, *f*. Fig. 8 shows the correction factor, *f*, for $k=0.3$ to 1.0, as a function of visibility size, as determined from the limb-darkening models discussed in the previous section. The correction factor should be unity when $k=0.5$ for a uniform disk, greater than unity for limb-darkened disk ($k>0.5$), and less than unity for limb-brightened disk ($k<0.5$). The correction factor not only depends on limb-darkening, but also the size of the object compared to fringe spacing. Fig. 8 shows that, for objects with diameters less than $\lambda/B_\perp$ the correction factors range from 0.95 ($k=0.3$) to 1.0 ($k=0.5$) and then to 1.14 ($k=1.0$).

When the object is much smaller than $\lambda/B_\perp$, the correction factor approaches a constant, which is proportional to the second moment of the brightness distribution, defined as $R_{2nd}^2 = \dfrac{\int r^2 B(r) d\Omega}{\int B(r) d\Omega}$, where $\Omega$ represents solid angle on the sky, and $B(r)$ is the surface brightness of the object at angle *r* from disk center. This is because that when the size approaches zero, the cosine kernel of the Fourier Transform approaches $r^2$.

Knowing the albedo of an object can usually put some constraints on its limb-darkening properties in the visible and/or near IR wavelengths. Usually, for dark surfaces, limb-darkening is weak at low phase angles even for high roughness (McEwen, 1991). E.g., the surface of the Moon has a geometric albedo of about 0.2 (Hillier et al., 1999), and an almost flat disk at low phase angle. For bright surfaces such as those covered by ices, the limb-darkening is relatively strong. The strongest limb-darkening, corresponding to a Minnaert *k* parameter of 1.0, is expected for a perfect scattering surface with an SSA of unity and isotropic scattering.

Although many KBOs have featureless spectra (Jewitt and Luu, 2001, Licandro et al., 2002), ices have been identified on several of them. The surface of Triton, which is possibly a captured object from the Kuiper Belt (Agnor and Hamilton, 2006), is covered by ices of $N_2$, $CH_4$, CO and $O_2$ (Brown et al., 1995), and has an SSA of 0.98 and moderate roughness of 15-20º (Hillier et al., 1994), resulting in a Minnaert *k* parameter of about 0.85. Both Eris and (90377) Sedna were identified to have surface compositions similar to that of Triton (Brown et al., 2005a;

Barucci et al., 2005). Water ices have been found on a handful of large KBOs, such as 1996 TO$_{66}$, 2004 DW, Quaoar, Haumea, etc. (Brown, et al., 1999; Fornasier et al., 2004; Jewitt and Luu, 2001; 2004, Trujillo et al., 2007). Therefore, the limb-darkening of large KBOs is expected to be strong, probably similar to that of Triton. The correction factors for the surfaces of large KBOs are therefore expected to range from 1.06-1.14, about a range of 8%. Although theoretically, limb-brightening ($k<0.5$) can exist for very rough and bright surfaces, it has never been observed in solar-system objects. Therefore, we conclude that the sizes of large, spherical KBOs measured by interferometry should have uncertainties less than about 8% even if their limb-darkening properties are completely unknown.

We can compare this analysis with the measurements of Quaoar's and Eris' sizes with HST. For Quaoar, Brown and Trujillo (2004) experimented with a range of limb-darkening models. They concluded that the apparent size is approximately proportional to the half-total light diameter, and derived a 15% uncertainty caused by the possible range of limb-darkening parameters. Brown et al. (2006) measured the size of Eris with HST and claimed 4% uncertainty, assuming that the surface of Eris is similar to that of Triton.

## 6. Surface brightness features, polar caps

Planetary bodies commonly have compositional or topographic surface features. Albedo features dominate the photometric light curves of Ceres, Vesta, and Pluto (Buie et al., 2010, and reference therein). The dark red spot (DRS) on Haumea is another interesting surface feature on a KBO (Lacerda et al., 2008). Brightness variations can indicate variations in composition, particle size, and regolith maturity, possibly associated with different local geological or collisional histories. E.g., the volcanoes on Io produce many time varying albedo variations on this body. The geologically active area near the south pole of Enceladus has albedo patterns (tiger stripes) resulted from different particle sizes released by active outgassing (Jaumann et al., 2008). The temporal change of albedo in some areas on the surface of Triton is possibly due to the seasonal N$_2$ ice evaporation and re-condensation cycle (Bauer et al., 2010). The drastically different albedos on the leading and trailing hemispheres of Iapetus represent coatings of materials from other Saturnian satellites, possibly including Phoebe (Soter, 1974) and Hyperion (Matthews, 1992). Measuring such features can yield important information about an object's history and environment, and also significantly improve our ability to determine an object's rotational state.

*6.1 The dark red spot on Haumea*

Taking the example of Haumea again with the DRS (Lacerda et al., 2008), we studied how surface brightness variations can be observed with interferometers. The thin lines in Fig. 4 shows the simulated photometric lightcurves and visibility lightcurves for Haumea with the DRS; Fig. 5b shows the phase center shift for an edge-on case; and the thin line in Fig. 6 for a pole-on case. The DRS is centered at 45º in longitude on Haumea (Lacerda et al., 2008). We arbitrarily chose a latitude of +15º for the center of the DRS. We assumed a radius of the DRS of 40º on the surface of the body, roughly 50% of the cross-sectional area when directly viewed, and an albedo that is 87% of the albedo of the rest of the surface, according to the observations by Lacerda et al. (2008). In these discussions, we assume that the proper motion of the object has been precisely extracted and removed. As shown in the upper left panel of Fig. 4, these assumptions reproduced the observed photometric lightcurves reported by Rabinowitz et al. (2006) and Lacerda et al. (2008).

Fig. 4 shows that adding in the DRS does not affect the range of the visibility lightcurves of the object. In fact, with the designed 1% accuracy to correlated flux for SIM, the thin lines and the thick lines in Fig. 4 are not distinguishable at all. For the nearly edge-on view, when the DRS swings into view, it could cause one minimum of the visibility lightcurve to shift by ~30º, which is probably detectable. Thus surface albedo features would not affect the derivation of the rotational period, and should not affect the ability to comparatively analyze the photometric lightcurves and visibility lightcurves for deriving the size and pole orientation of the objects.

On the other hand, adding the DRS affected the visibility phase dramatically. Without the albedo feature, the shape and limb-darkening of the surface determine the visibility phase, and the photocenter mostly traces the movement of the sub-solar point (usually the brightest point). However, the DRS dominates the displacement of photocenter, though it only changes the photometric light curves by 0.05 mag and does not change the range of visibility lightcurves.

As shown in Fig. 5b, while the DRS is visible, the phase center executes a loop about 200 µas across. When the DRS is not visible (rotational phase 0.2-0.55), the phase center moves in a loop of only about 80 µas across. Presented with the phase center displacement path and the photometric lightcurve, one could infer that this body had a dark spot rather than a bright spot because the total brightness of the body drops while the photocenter is executing the large loop. Once the bright or dark spot is determined, the track of the phase center would indicate both the sense of rotation and the projected orientation of pole.

*6.2 Polar caps*

Another kind of potentially important surface feature are polar caps. Polar caps are 1) usually bright, and 2) located near the rotational poles, so they do not move much in the course of an object's rotation. Therefore, polar caps do not contribute to lightcurves; they only cause a static change in the visibility, which we can identify if we have enough ($u$, $v$) coverage. A single unresolved polar cap would add a term $Ae^{i(ux+vy)(B_{\perp}/\lambda)}$ to the complex visibility, where ($x$, $y$) is the position of the cap measured with respect to the phase center; a pair of caps will add a cosine term. In interferometry, a cosine visibility pattern is generally associated with a binary object. But for a binary object, the cosine orientation and wavelength change as the binary rotates, while for polar caps, the cosine orientation will be fixed. The strength of the above modulation caused by polar caps, or parameter $A$ in above expression, will be the fraction of brightness from polar caps relative to the total brightness of the object and can be measured from visibilities. E.g., for Mars, the polar caps cover about a few percent of its disk. The albedo of the Mars desert is about 0.15, and the albedo of polar cap is about 0.85 (James et al., 2005), resulting in a parameter *A* of ~10%.

## 7. Considerations for SIM's applications in planetary sciences

We have discussed the applications of interferometry in planetary sciences in a general case and not tied to any specific instruments. In practice, specific instruments require specific considerations. Let us now discuss some operational details in applying the applications we discussed above that is specific for SIM.

*7.1 Operational consideration*

In order to measure the triaxial shape of a solar system small body, we need visibility measurements along at least three baseline orientations. If using SIM to observe, then we will have to rotate it to a new baseline orientation after each measurement. The time it takes for SIM to change baseline orientation (tens of minutes) is a considerable fraction of the rotational periods of many KBOs and Centaurs listed in Table 1. Since we generally have some knowledge about the rotational period of KBOs and Centaurs from ground-based radiometric observations, one can observe the target along one baseline orientation for at least a full rotation, then change to the next baseline and observe the target for another full rotation, and so on. As long as the various lightcurves can be phased correctly, we will be able to find visibility measurements from the same rotational phase of the target, and derive its shape. Since the observing geometry of KBOs changes in a time scale of years, and Centaurs in months at least, the observations along different baselines do not have to be scheduled next to each other, adding flexibility in the scheduling.

The brightness of an object, and thus the required exposure time, determines the time resolution of the various lightcurves discussed in previous sections. For the KBOs and Centaurs we considered in Table 1, their rotational periods range from 4 hrs to about 20 hrs. The integration time for a 20 mag star in visible wavelength required by SIM to reach 1% uncertainty in the correlated flux is about 1 hour, and longer for extended sources depending on its visibility. The time resolution of one to two hours is probably enough for a 20-hr period object. For a 4-hr period, it may have to be brighter than 19 mag in order to reach the same measurement precision. However, depending on the particular goal of observations, one can always perform on-chip binning over spectral channels to increase the signal and shorten the exposure time.

*7.2 Tracking of high-proper motion objects*

Solar system objects always have much higher, varying proper motions compared to stellar objects. For a typical KBO at 30 AU, the maximum proper motion near opposition caused by the relative motion between the Earth and the KBO will be about 1.2 mas/s, or one SIM fringe spacing per 15 seconds at 500 nm wavelength. The proper motion could be higher if the KBO is in an inclined orbit. The minimum possible proper motion of a planetary object occurs near quadratures, when only the component of the object's velocity perpendicular to the ecliptic plane contributes. For classical KBOs whose orbital inclinations are less than 10º, the maximum proper motion near quadrature can be as low as 0.2 mas/s. For the "dynamically hot" population, it can still be as high as 0.6 mas/s.

Current SIM designs include active tracking capabilities at up to 100 mas/s to compensate for telescope vibrations. If the rate of proper motion is known, it can be programmed into the delay line to actively compensate for the phase change of targets introduced by proper motion. This rate should be sufficient to track most solar system objects in most cases. For objects brighter than 12 mag, SIM performs fringe measurement in milli-sec. In this case, the active tracking does not need to be utilized at all, and SIM will measure the phase center change, which includes proper motion, at a frequency of several hundred Hz. For relatively faint but fast moving objects, the integration time required to reach 1% accuracy would lead to smearing the data. The proper motion, which has usually been measured to a considerable precision for most KBOs and Centaurs, will have to be programmed in. This allows an enormous increase in the permissible integration time. If the proper motion rate is slightly wrong, then the fringe phase will slowly drift away. But this secondary effect is included in the data stream and can be measured and analyzed on the ground.

Asteroids move much faster than KBOs. At 3 AU, the maximum proper motion of an object in a prograde orbit in the ecliptic plane is about 9 mas/s, or one SIM fringe per 2 seconds at 500 nm wavelength. This is still within the capability of SIM. On the other hand, the asteroids whose sizes match the fringe spacing of SIM are much brighter than KBOs, and their proper motion should be less a problem than KBOS.

*7.3 Using the 4.2-m guide interferometer for sciences*

The largest KBOs, such as Eris, MakeMake, Haumea, and Quaoar are marginally too large to be measured by SIM 6-m interferometer (Fig. 1). The measurement might be unreliable and could have large uncertainties because of the low visibilities. These targets would be better served by SIM's 4.2-m guide interferometer. With its larger fringe spacing, the guide interferometer could measure accurate visibility for objects that are about 50% larger, covering even more objects that can be measured by the science interferometer in the objects that we considered (dark-gray shade in Fig. 1). The science interferometer can serve as the guide for these measurements.

Assuming there is no change in the brightness distribution of a target, if observing with the 4.2-m guide interferometer at 900 nm wavelength, then the fringe spacing is equivalent to the diffraction limit of HST at 450 nm wavelength. This essentially provides a bridge to combine

HST images with SIM observations.  For the objects such as Eris and Quaoar that have been observed by HST, if they can be observed by SIM with both science and guide interferometers from 400 nm to 900 nm, then the ($u$, $v$) coverage of those objects would be substantially enriched, and overlapped with HST.  The visibility data collected in the shortest wavelength of 450 nm with the 6-m science interferometer would provide spatial resolution that is about 1/3 of that of HST.  If enough ($u$, $v$) coverage can be obtained, then it is even possible to construct a surface albedo maps for those objects, although it could be extremely time-consuming.

The guide interferometer on SIM provides us with much more capabilities than guiding.  It substantially enlarges the ($u$, $v$) coverage of the science interfereometer, bridges SIM to other telescopes such as HST, as well as provides redundancy to the science interferometer.

## 8. Conclusions.

To summarize, we considered the use of interferometry to measure the basic physical properties of solar system small bodies, including their sizes, shapes, and rotations.  Those fundamental physical properties determine the conditions on their surfaces and interior, and contain information about the formation and evolution of the solar system.  Because of their small angular sizes, most of those objects cannot be spatially resolved and directly imaged with current single-dish telescopes.  Their fundamental properties are almost exclusively measured through indirect methods, which are generally model-dependent and have many limitations from observing geometry.  Current size measurements are typically uncertain at the 20% level, which is not enough to reliably derive many fundamental properties including densities, surface gravity, and internal structure.

Taking SIM as our primary example of an interferometer, we examined about 140 KBOs and Centaurs and identified 38 potentially scientifically interesting objects that are observable by SIM.  SIM should be able to observe objects up to 100 AU away from the Sun with diameters between 2000 and 3000 km.  It can observe objects at 10 AU from the Sun with diameters between 20 and 200 km, and objects down to 2 km in diameter in the Main Asteroid Belt.  ALMA, when operated in its most extended configuration, has similar sensitivity and spatial resolution as SIM in most regions from Main Asteroid Belt to Kuiper Belt.  But SIM has slightly higher sensitivity than ALMA for KBOs.  Because they operate in entirely different spectral ranges, together SIM and ALMA possess complementary capabilities in observing solar system small bodies.

We showed that, assuming an ellipsoidal disk model, the projected size and shape of an object could be derived from visibility measurements given a minimum of three different interferometer baseline orientations.  We derived a set of equations for converting visibility measurements to parameters of this model for two sets of baseline orientations, 45º separations and 60º separations.  The best measurement occurs when one axis of the elliptical disk is aligned with a baseline, and the uncertainty of the size measurement would be comparable to visibility measurement, assuming no limb-darkening. The cross-sectional area of the disk can always be measured to a comparable uncertainty of visibility regardless of the baseline orientations.

We constructed a 3-D shape model of the dwarf planet Haumea and used it to synthesize interferometric data, including the object's rotation. The rotational period can be derived using the techniques similar to those applied to analyze photometric lightcurves. Since the visibility variation of an object is determined by the variation of object's size along the corresponding interferometer baseline, while photometric variation is determined by the cross-sectional area, the combination of both enables us to derive the true axial ratio and the orientation of rotational pole.

The visibility phase center indicates the displacement of an object's photocenter relative to its geometric center. Without surface features, the visibility phase center closely tracks the brightest (approximately sub-solar) point when the object has an ellipsoidal shape. For our models of Haumea, the phase center describes a loop roughly 80 µas across as the object rotates, because of Haumea's elongated shape and limb darkening, without Haumea's dark red spot. When we added the spot, we found that the spot caused <5% change in the total brightness, but a large phase center displacement during the object's rotation: ~200 µas. That displacement can be easily measured by SIM. The path of an object's phase center, when combined with photometric lightcurves and visibility amplitude lightcurves, can also be used to determine the rotational state of an asteroid.

The non-uniform brightness distributions, usually limb-darkening, of solar system small bodies affect the accuracy of their size measurements. We explored various limb-darkening parameters, and demonstrated that assuming the Minnaert limb-darkening model, the limb-darkening parameter can be determined from a set of visibilities simultaneously obtained at multiple $\lambda/B_\perp$. This determination further improves the accuracy of an interferometric size measurement. However, even with the limb-darkening properties entirely unknown, one can still measure an object's size to 8% uncertainty through interferomety. We calculated the correction factors for a range of limb-darkening parameters to the size inferred from visibility, assuming a uniform disk, for use in this case. With correction for limb-darkening, interferometric techniques can generally measure the size of a solar system small bodies to an accuracy of 5%.

We find that a space-based interferometer, such as SIM with the current design characteristics can afford most KBO and asteroid sciences discussed here with some considerations according to the way it is operated. Its active tracking capability can effectively track high-proper motion solar system objects. The availability of the 4.2-m guide interferometer for science purposes will cover most of the object that can be observed by the science interferometer, and enable SIM to observe objects that are 50% larger, and substantially enrich the $(u, v)$ coverage that is especially valuable for solar system small bodies.

**Acknowledgements**

We thank the NASA Exoplanet Science Institute for support of this research via a SIM Science Studies grant. The authors are extremely grateful to the reviewers, who provided reviews that have helped improve this manuscript substantially.
**References**

Agnor, C.B., Hamilton, D.P., 2006. Neptune's capture of its moon Triton in a binary-planet gravitational encounter. Nature 441, 192-194.

Barucci, M.A., Cruikshank, D.P., Dotto, E., Merlin, F., Poulet, F., Dalle, O.C., Fornasier, S., de Bergh, C., 2005. Is Sedna another Triton? Astron. Astrophys. 439, 1-4.

Bauer, J., Meech, K., Fernandez, Y., Farnham, T., Roush, T., 2002. Observations of the Centaur 1999 $UG_5$: Evidence of a unique outer solar system surface. Publ. Astron. Soc. Pac. 114, 1309–1321.

Bauer, J., Meech, K., Fernandez, Y., Pittichova, J., Hainaut, O., Boehnhardt, H., Delsanti, A., 2003. Physical survey of 24 Centaurs with visible photometry. Icarus 166, 195–211.

Bauer, J.M., Buratti, B.J., Li, J.-Y., Mosher, J.A., Hicks, M.D., Schmidt, B.E., Goguen, J.D., 2010. Direct detection of seasonal changes on Triton with HST. Astrophys. J. Lett 723, 49-52.

Bernstein, G., Trilling, D., Allen, R., Brown, M., Holman, M., Malhotra, R., 2004. The size distribution of trans-Neptunian bodies. Astron. J. 128, 1364-1390.

Binzel, R.P., Gaffey, M.J., Thomas, P.C., Zellner, B.H., Storrs, A.D., Wells, E.N., 1997. Geologic mapping of Vesta from 1994 Hubble Space Telescope images. Icarus 128, 95-103.

Brown, R.H., Cruikshank, D.P., Veverka, J., Helfenstein, P., Eluszkiewicz, J., 1995. Surface composition and photometric properties of Triton. In: Cruikshank, D.P. (Ed.), Neptune and Triton. Univ. of Arizona Press, Tucson, pp. 991–1030.

Brown, R.H., Cruikshank, D.P., Pendleton, Y., 1999. Water ice on Kuiper Belt object 1996 TO66. Astrophys. J. Lett. 519, 101-104.

Brown, M.E., Trujillo, C.A., 2004. Direct measurement of the size of the large Kuiper Belt object (50000) Quaoar. Astron. J. 127, 2413-2417.

Brown, M.E., Trujillo, C.A., Rabinowitz, D.L., 2005a. Discovery of a planetary-sized object in the scattered Kuiper Belt. Astrophys. J. 635, 97-100.
26

Brown, M.E., Schaller, E.L., Roe, H.G., Rabinowitz, D.L., Trujillo, C.A., 2006. Direct measurement of the size of 2003 $UB_{313}$ from the Hubble Space Telescope. Astrophys. J. 643, L61-L63.

Brown, W., Luu, J., 1997. CCD photometry of the Centaur 1995 GO. Icarus 126, 218–224.

Buie, M., Bus, S., 1992. Physical observations of (5145) Pholus. Icarus 100, 288–294.

Buie, M.W., Tholen, D.J., Wasserman, L.H., 1997. Separate lightcurves of Pluto and Charon. Icarus 125, 233-244.

Buie, M.W., Grundy, W.M., Young, E.F., Young, L.A., Stern, S.A., 2010. Pluto and Charon with the Hubble Space Telescope. II. Resolving changes on Pluto's surface and a map for Charon. Astron. J. 139, 1128-1143.

Bus, S., Bowell, E., Harris, A., Hewitt, A., 1989. 2060 Chiron – CCD and electronographic photometry. Icarus 77, 223–238.

Busch, M.W., 2009. ALMA and asteroid science. Icarus 200, 347-349.

Castillo-Rogez, J.C., McCord, T.B., 2009. Ceres' evolution and present state constrained by shape data. Icarus 205, 443-459.

Chamberlain, M.A., Sykes, M.V., Esquerdo, G.A., 2007. Ceres lightcurve analysis – period determination. Icarus 188, 451-456.

Davies, J., McBride, N., Ellison, S., Green, S., Ballantyne, D., 1998. Visible and infrared photometry of six Centaurs. Icarus 134, 213–227.

Davis, D., Farinella, P., 1997. Collisional evolution of Edgeworth-Kuiper belt objects. Icarus 125, 50-60.

Davis, D.R., Durda, D.D., Marzari, F., Campo Bagatin, A., Gil-Hutton, R., 2002. Collisional evolution of small-body populations. In: Bottke, W.F., Cellino, A., Paolicchi, P., Binzel, R.P. (Eds.), Asteroids III. University of Arizona Press, Tucson, pp. 545-558.

Davidson, J., Edberg, S., Danner, R., Nemati, B., Unwin, S., (Eds.) 2009. JPL Pub. No. 400-1360 dd. 2/09.

Delbo, M., Harris, A.W., 2002. Physical properties of near-Earth asteroids from thermal infrared observations and thermal modeling. Meteorit. Planet. Sci. 37, 1929-1936.

Delbo, M., Ligori, S., Matter, A., Cellino, A., Berthier, J., 2009. First VLTI-MIDI direct determinations of asteroid sizes. Astrophys. J. 694, 1228-1236.

Farnham, T., 2001. The rotation axis of Centaur 5145 Pholus. Icarus 152, 238–245.

Farinella, P., Davis, D., 1996. Short-period comets: Primordial bodies or collisional fragments? Science 273, 938-941.

Fornasier, S., Dotto, E., Barucci, M.A., Barbieri, C., 2004. Water ice on the surface of the large TNO 2004 DW. Astron. Astrophys. 422, 43-46.

Gendzwill, D.J., Stauffer, M.R., 1981. Analysis of triaxial ellipsoids: their shapes, plane sections, and plane projections. Mathematical Geology 13, 135-152.

Gutierrez, P., Oritz, J., Alexandrino, E., Roos-Serote, M., Doressoundiram, A., 2001. Short term variability of Centaur 1999 $UG_5$. Astron. Astrophys. 371, L1–L4.

Hajian, A.R., and 10 colleagues, 1998. Direct confirmation of stellar limb darkening with the Navy prototype optical interferometer. Astrophys. J. 496, 484-489.

Hapke, B., 1993. Theory of reflectance and emittance spectroscopy. Cambridge Univ. Press, Cambridge, UK.

Harris, A.W., 1998. A thermal model for near-Earth asteroids. Icarus 131, 291-301.

Harris, A.W., Lagerros, J.S.V., 2002. Asteroids in the thermal infrared. In: Bottke, W.F., Cellino, A., Paolicchi, P., Binzel, R.P., (Eds.), Asteroids III, Univ. of Arizona Press, Tucson, pp. 205–218.

Hestroffer, D., Tanga, P., Cellino, A., Guglielmetti, F., Lattanzi, M., Di Martino, M., Zappalà, V., Berthier, J., 2002. Asteroids observations with the Hubble Space Telescope FGS I. Observing strategy, and data analysis and modeling process. Astron. Astrophys. 391, 1123-1132.

Hillier, J., Veverka, J., Helfenstein, P., Lee, P., 1994. Photometric diversity of terrains on Triton. Icarus 109, 296-312.

Hillier, J.K., Buratti, B.J., Hill, K., 1999. Multispectral photometry of the Moon and absolute calibration of the Clementine UV/Vis camera. Icarus 141, 205-225.

Hoffmann, M., Fink, U., Grundy, W., Hicks, M., 1993. Photometric and spectroscopic observations of 5145 Pholus. J. Geophys. Res. 98, 7403–7407.

Horner, J., Evans, N.W., Bailey, M.E., 2004. Simulations of the population of Centaurs – I. The bulk statistics. Mon. Not. Roy. Astron. Soc. 354, 798-810.

James, P.B., Bonev, B.P., Wolff, M.J., 2005. Visible albedo of Mars' south polar cap: 2003 HST observations. Icarus 174, 596-599.

Jaumann, R., and 18 colleagues, 2008. Distribution of icy particles across Enceladus' surface as derived from Cassini-VIMS measurements. Icarus 193, 407-419.

Jewitt, D.C., Luu, J.X., 2001. Colors and spectra of Kuiper Belt Objects. Astron. J. 122, 2099-2114.

Jewitt, D.C., Sheppard, S.S., 2002. Physical properties of Trans-Neptunian object (20000) Varuna. Astron. J. 123, 2110-2120.

Jewitt, D.C., Luu, J., 2004. Crystalline water ice on the Kuiper belt object (50000) Quaoar. Nature 432, 731-733.

Johnstone, D., Di Francesco, J., Plume, R., Schieven, G.H.M., 2010. Observing with ALMA, a primer. http://almatelescope.ca/ALMAPrimer.pdf.

Kaasalainen, M., Mottola, S., Fulchignoni, M., 2002. Asteroid models from disk-integrated data. In: Bottke, W.F., Cellino, A., Paolicchi, P., Binzel, R.P., (Eds.), Asteroids III, Univ. of Arizona Press, Tucson, pp. 139-150.

Lacerda, P., Luu, J., 2006. Analysis of the rotational properties of Kuiper Belt objects. Astron. J. 131, 2314–2326.

Lacerda, P., Jewitt, D.C., 2007. Densities of solar system objects from their rotational light curves. Astron. J. 133, 1393-1408.

Lacerda, P., Jewitt, D., Peixinho, N., 2008. High-precision photometry of extreme KBO 2003 $EL_{61}$. Astron. J. 135, 1749-1756.

Lebofsky, L.A., Sykes, M.V., Tedesco, E.F., Veeder, G.J., Matson, D.L., Brown, R.H., Gradie, J.C., Feierberg, M.A., Rudy, R.J., 1986. A refined "standard" thermal model for asteroids based on observations of 1 Ceres and 2 Pallas. Icarus 68, 239-251

Levison, H.F., Duncan, M.J., 1997. From the Kuiper Belt to Jupiter-Family comets: the spatial distribution of ecliptic comets. Icarus 127, 13-32.

Li, J.-Y., A'Hearn, M.F., McFadden, L.A., 2004. Photometric analysis of Eros from NEAR data. Icarus 172, 415-431.

Li, J.-Y., McFadden, L.A., Parker, J.Wm., Young, E.F., Stern, A.S., Thomas, P.C., Russell, C.T., Sykes, M.V., 2006. Photometric analysis of 1 Ceres and surface mapping from HST observations. Icarus 182, 143-160.

Li, J.-Y., and 10 colleagues, 2007a. Deep Impact photometry of comet 9P/Tempel 1. Icarus 187, 41-55.

Li, J.-Y., A'Hearn, M.F., McFadden, L.A., Belton, M.J.S., 2007b. Photometric analysis and disk-resolved thermal modeling of comet 19P/Borrelly from Deep Space 1 data. Icarus 188, 195-211.

Li, J.-Y., A'Hearn, M.F., Farnham, T.L., McFadden, L.A., 2009. Photometric analysis of the nucleus of comet 81P/Wild 2 from Stardust images. Icarus 204, 209-226.

Li, J.-Y., McFadden, L.A., Thomas, P.C., Mutchler, M.J., Parker, J.Wm., Young, E.F., Russell, C.T., Sykes, M.V., Schmidt, B.E., 2010. Photometry and surface mapping of asteroid (4) Vesta's southern hemisphere with Hubble Space Telescope. Icarus 208, 238-251.

Licandro, J., Ghinassi, F., Testi, L., 2002. Infrared spectroscopy of the largest known trans-Neptunian object 2001 KX$_{76}$. Astron. Astrophys. 388, 9-12.

Lovell, A.J., 2008. Observatiosn of asteroids with ALMA. Astrophys. Space Sci. 313, 191-196.

Luu, J., Jewitt, D., 1990. Cometary activity in 2060 Chiron. Astron. J. 100, 913–932.

Marcialis, R., Buratti, B., 1993. CCD photometry of 2060 Chiron in 1985 and 1991. Icarus 104, 234–243.

Matthews, R.A.J., 1992. The darkening of Iapetus and the origin of Hyperion. Q. J. R. Astron. Soc. 33, 253-258.

McCord, T.B., Sotin, C., 2005. Ceres: Evolution and current state. J. Geophys. Res. 110, doi:10.1029/2004JE002244. E05009.

McEwen, A.S., 1991. Photometric functions for photoclinometry and other applications. Icarus 92, 298-311.

Michalak, G., 2000. Determination of asteroid masses – I. (1) Ceres, (2) Pallas, and (4) Vesta. Astron. Astrophys. 360, 363-374.

Minnaert, M., 1941. The reciprocity principle in lunar photometry, Astrophys. J. 93, 403-410.

Mueller, B., Hergenrother, C., Samarasinha, N., Campins, H., McCarthy, D., 2004. Simultaneous visible and near-infrared time resolved observations of the outer solar system object (29981) 1999 TD$_{10}$. Icarus 171, 506–515.

Ortiz, J., Baumont, S., Gutierrez, P., Roos-Serote, M., 2002. Lightcurves of Centaurs 2000 QC$_{243}$ and 2001 PT$_{13}$. Astron. Astrophys. 388, 661–666.

Ortiz, J., Gutierrez, P., Casanova, V., Sota, A., 2003a. A study of short term variability in TNOs and Centaurs from Sierra Nevada observatory. Astron. Astrophys. 407, 1149–1155.

Ortiz, J., Gutierrez, P., Sota, A., Casanova, V., Teixeira, V., 2003b. Rotational brightness variations in trans-Neptunian object 50000 Quaoar. Astron. Astrophys. 409, L13–L16.

Ortiz, J., and 10 colleagues, 2004. A study of trans-Neptunian object 55636 (2002 TX$_{300}$). Astron. Astrophys. 420, 383–388.

Ortiz, J.L., Gutiérrez, P.J., Santos-Sanz, P., Casanova, V., Sota, A., 2006. Short-term rotational variability of eight KBOs from Sierra Nevada Observatory. Astron. Astrophys. 447, 1131-1144.

Pravec, P., Harris, A.W., Michalowski, T., 2002. Asteroid Rotations. In: Bottke, W.F., Cellino, A., Paolicchi, P., Binzel, R.P. (Eds.), Asteroids III. University of Arizona Press, Tucson, pp. 113-122.

Quirrenbach, A., Mozurkewich, D., Buscher, D.F., Hummel, C.A., Armstrong, J.T., 1996. Angular diameter and limb darkening of Arcturus. Astron. Astrophys. 312, 160-166.

Rabinowitz, D.L., Barkume, K., Brown, M.E., Roe, H., Schwartz, M., Tourtellotte, S., Trujillo, C., 2006. Photometric observations constraining the size, shape, and albedo of 2003 EL61, a rapidly rotating, Pluto-sized object in the Kuiper belt. Astrophys. J. 639, 1238-1251.

Romanishin, W., Tegler, S.C., 2007. KBO and Centaur Absolute Magnitudes V1.0. EAR-A-VARGBDET-3-KBOMAGS-V1.0. NASA Planetary Data System, 2007.

Rousselot, P., Petit, J., Poulet, F., Lacerda, P., Ortiz, J., 2003. Astron. Astrophys. 407, 1139–1147.

Rousselot, P., Petit, J., Poulet, F., Sergeev, A., 2005. Photometric study of Centaur (60558) 2000 $EC_{98}$ and trans-Neptunian object (55637) 2002 $UX_{25}$ at different phase angles. Icarus 176, 478–491.

Schaefer, B., Rabinowitz, D., 2002. Photometric light curve for the Kuiper belt object 2000 $EB_{173}$ on 78 nights. Icarus 160, 52–58.

Schaller, E.L., Brown, M.E., 2007. Detection of methane on Kuiper belt object (50000) Quaoar. Astrophys. J. 670, 49-51.

Shepard, M.K., Cambell, B.A., 1998. Shadows on a planetary surface and implications for photometric roughness. Icarus 134, 279-291.

Shepard, M., Helfenstein, P., 2007. A test of the Hapke photometric model. J. Geophys. Res. 112, 1-17. E03001.

Sheppard, S., Jewitt, D., 2002. Time-resolved photometry of Kuiper belt objects: Rotations, shapes, and phase functions. Astron. J. 124, 1757–1775.

Sheppard, S. Jewitt, D., 2003. Hawaii Kuiper belt variability project: An update. Earth Moon Planets 92, 207–219.

Sheppard, S.S., Jewitt, D.C., 2004. Extreme Kuiper Belt object 2001 $QG_{298}$ and the fraction of contact binaries. Astron. J. 127, 3023-3033.

Sheppard, S., 2007. Light curves of dwarf plutonian planets and other large Kuiper belt objects: Their rotations, phase functions, and absolute magnitudes. Astron. J. 134, 787–798.

Sheppard, S.S., Lacerda, P., Ortiz, J.L., 2008. Photometric lightcurves of transneptunian objects and Centaurs: rotations, shapes, and densities. In: Barucci, M.A., Boehnhardt, H., Cruikshank, D.P., Morbidelli, A. (Eds.), The solar system beyond Neptune. Univ. of Arizona Press, Tucson, pp. 129-142.

Sivaramakrishnan, A., and 9 colleagues, 2009. Planetary system and star formation science with non-redundant masking on JWST. In: Shaklan, S.B. (Ed), Techniques and Instrumentation for Detecion of Exoplanets IV. Proc. SPIE, 7440, 30.

Soter, S., 1974. Brightness of Iapetus. Presented at IAU Colloq. 28, Cornell Univ.

Stansberry, J., Grundy, W., Brown, M., Cruikshank, D., Spencer, J., Trilling, D., Margot, J.-L., 2008. Physical properties of Kuiper Belt and Centaur objects: Constraints from Spitzer Space Telescope. In: Barucci, M.A., Boehnhardt, H., Cruikshank, D.P., Morbidelli, A. (Eds.), The solar system beyond Neptune. Univ. of Arizona Press, Tucson, pp. 161-179.

Stellingwerf, R.F., 1978. Period determination using phase dispersion minimization. Astrophys. J. 224, 953-960.

Takahashi, S., Ip, W.-H., 2004. A shape-and-density model of the putative binary EKBO 2001 $QG_{298}$. Publ. Astron. Soc. Jpn. 56, 1099-1103.

Tanga, P., Hestroffer, D., Cellino, A., Lattanzi, M., Di Martino, M., Zappalà, V., 2003. Asteroid observations with the Hubble Space Telescope. II. Duplicity search and size measurements for 6 asteroids. Astron. Astrophys. 401, 733-741.

Tanga, P., Comito, C., Paolicchi, P., Hestroffer, D., Cellino, A., Dell'Oro, A., Richardson, D.C., Walsh, K.J., Delbo, M., 2009. Rubble-pile reshaping reproduces overall asteroid shapes. Astrophys. J. Lett. 706, 197-202.

Tarenghi, M., 2008. The Atacama Large Millimeter/Submillimeter Array: overview & status. Astrophys. Space Sci 313, 1-7.

Tegler, S., Romanishin, W., Consolmagno, G., Rall, J., Worhatch, R., Nelson, M., Weidenschilling, S., 2005. The period of rotation, shape, density, and homogeneous surface color of the Centaur 5145 Pholus. Icarus 175, 390–396.

Thomas, P.C., Veverka, J., Simonelli, D., Helfenstein, P., Carcich, B., Belton, M.J.S., Davies, M.E., Chapman, C., 1994. The shape of Gaspra. Icarus, 107, 23

Trujillo, C.A., Brown, M.E., Barkume, K.M., Schaller, E.L., Rabinowitz, 2007. The surface of 2003 $EL_{61}$ in the near-infrared. Astrophys. J. 655, 1172-1178.

Unwin, S.C., and 35 colleagues, 2008. Taking the measure of the universe: Precision astrometry with SIM PlanetQuest. PASP 120, 38-88.

Viateau, B., Rapport, N., 2001. Mass and density of asteroids (4) Vesta and (11) Parthenope. Astron. Astrophys. 370, 602-609.

**Table 1. Candidate KBOs and Centaurs for observation with SIM**

Columns:
    q: Perihelion distance
    H: Absolute magnitude
    V: Apparent magnitude near perihelion
    delta_m: Lightcurve magnitude
    Periods: Rotational periods fitted from lightcurves assuming single-peaked (s) or double-peaked (d)
    D: Diameter
    Ang.D: angular diameter near perihelion
    Visibility: Interferometric visibility assuming 15 mas fringe spacing (corresponding to SIM 6-m
        baseline operating at 450 nm wavelength)
* Diameter estimated by assuming an albedo (see text)

Reference code:

BB: Buie and Bus, 1992; BL: Brown and Luu, 1997; BM: Bauer et al., 2002; BMF: Bauer et al., 2003; BT: Brown and Trujillo, 2004; BBH: Bus et al., 1989; DN: Davies et al., 1998; F: Farnham, 2001; GO: Gutierrez et al., 2001; H: Hoffmann et al., 1993; JS: Jewitt and Sheppard, 2002; L: Luu and Jewitt, 1990; LL: Lacerda and Luu, 2006; MB: Marcialis and Buratti, 1993; MH: Mueller et al., 2004; OBG: Ortiz et al., 2002; OGC: Ortiz et al., 2003a; OG3: Ortiz et al., 2003b; OSM: Ortiz et al., 2004; OGS: Ortiz et al., 2006; PDS: Romanishin and Tegler, 2007; RBB: Rabinowitz et al., 2006; RP5: Rousselot et al., 2005b; RRP: Rousselot et al., 2003; SR: Schaefer and Rabinowitz, 2002; SJ2: Sheppard and Jewitt, 2002; SJ3: Sheppard and Jewitt, 2003; S: Sheppard, 2007; SGB: Stansberry et al., 2008; TRC: Tegler et al., 2005.

| Number | Name | q (AU) | H (mag) | V (mag) | delta_m (mag) | Periods (hrs) (hrs) | D (km) | Ang.D (mas) | Visibility | Ref |
|---|---|---|---|---|---|---|---|---|---|---|
| | **KBOs** | | | | | | | | | |
| 136108 | *Haumea (2003 EL 61) | 35.10 | 0.1 | 15.5 | 0.28+/-0.04 | 3.9154+/-0.0002 (d) | *1663 | 65.33 | 0.011 | RBB |
| 90482 | *Orcus (2004 DW) | 30.70 | 2.3 | 17.1 | 0.04+/-0.02 | 10.08+/-0.01 (s) | 946.3 | 42.50 | 0.058 | OGS |
| 50000 | *Quaoar (2002 LM 60) | 41.83 | 2.6 | 18.8 | 0.03+/-0.03 | 17.6788+/-0.0004 (d) | 1260 | 41.53 | 0.062 | OG3, BT |
| 28978 | Ixion (2001 KX 76) | 30.13 | 3.2 | 17.9 | <0.05 | | 573.1 | 26.22 | -0.124 | SJ3, OGC |
| 55565 | (2002 AW 197) | 41.19 | 3.3 | 19.4 | 0.08+/-0.07 | 8.86+/-0.01 (s) | 734.6 | 24.59 | -0.132 | OGS |
| 55636 | (2002 TX300) | 37.93 | 3.3 | 19.0 | 0.08+/-0.02 | 8.0, 12.1 (s), 16.0, 24.2 (d) | 641.2 | 23.31 | -0.128 | SJ3, OSM |
| 55637 | *(2002 UX 25) | 36.57 | 3.6 | 19.2 | <0.06, 0.21+/-0.06 | 7.2, 8.4 (s) | 681.2 | 25.68 | -0.129 | RP5 |
| 208996 | *(2003 AZ 84) | 32.96 | 3.9 | 19.0 | 0.12+/-0.02 | 6.72 (s), 13.44 (d) | 685.8 | 28.69 | -0.092 | SJ3, OGS |
| 20000 | Varuna (2000 WR 106) | 40.77 | 3.9 | 20.0 | 0.42+/-0.03 | 6.3436+/-0.0002 | 621.2 | 21.01 | -0.092 | F, SJ2, JS, OGC, PDS |
| | (2002 MS 4) | 35.33 | 4.0 | 19.4 | | | 726.2 | 28.34 | -0.097 | PDS |
| 90568 | (2004 GV 9) | 38.86 | 4.0 | 19.8 | <0.08 | | 677.2 | 24.03 | -0.132 | S |
| 84522 | (2002 TC 302) | 39.07 | 4.1 | 20.0 | | | 1145.4 | 40.42 | 0.064 | PDS |
| 42301 | (2001 UR 163) | 37.26 | 4.2 | 19.9 | <0.08 | | *617 | 22.81 | -0.123 | SJ3 |
| 84922 | (2003 VS 2) | 36.52 | 4.2 | 19.8 | 0.21+/-0.02 | 7.41 (d) | 725.2 | 27.38 | -0.112 | S, OGS |
| 120348 | (2004 TY 364) | 36.50 | 4.5 | 20.1 | 0.22+/-0.02 | 5.85+/-0.01 (s), 11.70+/-0.01 (d) | *537 | 20.28 | -0.073 | S |
| 26375 | (1999 DE9) | 32.35 | 4.7 | 19.7 | <0.10 | >12? (s) | 461 | 19.65 | -0.053 | SJ2 |
| | *(2001 QF 298) | 35.27 | 4.7 | 20.1 | <0.12 | | *490 | 19.15 | -0.035 | SJ3 |
| 38628 | Huya (2000 EB 173) | 28.64 | 4.7 | 19.2 | <0.06 | | 546.5 | 26.31 | -0.124 | SJ2, SR, OGC, LL |
| 47171 | *(1999 TC36) | 30.62 | 4.9 | 19.7 | <0.05 | | 414.6 | 18.67 | -0.016 | SJ3, OGC, LL |
| 55638 | (2002 VE 95) | 27.98 | 5.3 | 19.7 | <0.06, 0.08+/-0.04 | 6.76, 7.36, 9.47 (s) | *372 | 18.31 | 0.000 | SJ3, OGS |
| | **Centaurs** | | | | | | | | | |
| 2060 | Chiron (1977 UB) | 8.45 | 6.6 | 15.6 | 0.09 to 0.45 | 5.917813 (d) | 233.3 | 38.05 | 0.058 | BBH, L, MB, SGB |
| 65489 | Ceto (2003 FX 128) | 17.49 | 6.6 | 18.9 | | | 229.7 | 18.11 | 0.008 | PDS |
| 10199 | Chariklo (1997 CU 26) | 13.07 | 6.8 | 17.8 | | | 260.9 | 27.51 | -0.110 | PDS |
| 5145 | Pholus (1992 AD) | 8.78 | 7.0 | 16.2 | 0.15 to 0.6 | 9.98 (d) | 138.9 | 21.80 | -0.109 | BB, H, F, TRC |
| 42355 | Typhon (2002 CR 46) | 17.53 | 7.2 | 19.5 | <0.05 | | *155 | 12.18 | 0.380 | SJ3, OGC |
| 95626 | (2002 GZ 32) | 18.07 | 7.2 | 19.7 | | | *152 | 11.60 | 0.423 | PDS |
| 54598 | Bienor (2000 QC 243) | 13.18 | 7.6 | 18.6 | 0.75+/-0.09 | 4.57+/-0.02 (s) | 206.7 | 21.63 | -0.105 | OBG |
| 55576 | Amycus (2002 GB 10) | 15.16 | 8.1 | 19.7 | | | 76.3 | 6.94 | 0.758 | PDS |
| 73480 | (2002 PN 34) | 13.32 | 8.2 | 19.3 | 0.18+/-0.04 | 4.23, 5.11 (s) | 119.5 | 12.37 | 0.365 | OGC |
| 29981 | (1999 TD 10) | 12.36 | 8.8 | 19.5 | 0.65+/-0.05 | 7.71+/-0.02 (s) | 103.7 | 11.57 | 0.425 | OGC, RPP, MH |
| 120061 | (2003 CO 1) | 10.90 | 8.9 | 19.1 | 0.10+/-0.05 | 4.99 (s) | 76.9 | 9.73 | 0.563 | OGS |
| 8405 | Asbolus (1995 GO) | 6.83 | 9.2 | 17.2 | 0.55 | 8.93 (d) | 83.2 | 16.78 | 0.075 | BL, DN |
| 32532 | Thereus (2001 PT13) | 8.49 | 9.3 | 18.3 | 0.16+/-0.02 | 4.1546+/-0.0001 (s) | 60.8 | 9.87 | 0.553 | OBG, PDS |
| 60558 | Echeclus (2000 EC 98) | 5.82 | 9.5 | 16.7 | 0.24+/-0.06 | 13.401 (s) | 83.6 | 19.82 | -0.058 | RP5 |
| 7066 | Nessus (1993 HA 2) | 11.82 | 9.5 | 20.1 | | | *53 | 6.21 | 0.803 | PDS |
| 31824 | Elatus (1999 UG5) | 7.30 | 10.1 | 18.4 | 0.102 to 0.24 | 13.25, 13.41 (s) | 34.1 | 6.44 | 0.789 | BM, GO |
| 52872 | Okyrhoe (1998 SG 35) | 5.79 | 11.3 | 18.5 | 0.2 | 8.3 | 52.1 | 12.41 | 0.362 | BMF |
| 63252 | (2001 BL 41) | 6.88 | 11.5 | 19.5 | | | *22 | 4.36 | 0.899 | PDS |

**Figure captions**

Fig. 1 - KBOs and Centaurs observable with SIM and ALMA. The symbols are 109 KBOs and 26 Centaurs with sizes estimated from IR radiometry (Stansberry et al., 2008) or from their absolute magnitudes, *H*, with assumed geometric albedo of 0.1 for objects smaller than 600 km in diameter and 0.6 for objects larger than 600 km. The uncertainties of the size measurements are typically 20% or more. The stars mark a subset of 21 KBOs and 17 Centaurs (listed in Tabel 1) that SIM should be able to observe. The color, shade, and lines are: (a) the yellow shaded area is where SIM 6-m science interferometer can observe assuming a geometric albedo of 0.2; (b) the blue shaded area is where SIM 4.2-m guide interferometer can observe; (c) the green shaded area is where both SIM 6-m science and 4.2-m guide interferometers can observe; (d) the dotted area is observable by ALMA, assuming 1.3 mJy continuum sensitivity at 675 GHz; (e) the solid curves mark the objects that subtend constant angular diameters as noted on the right side of the plot in mas; (f) the dashed lines correspond to *V*=20 magnitude, SIM's limiting magnitude, with assumed geometric albedos from 0.02 to 1.0; and (g) the dotted curves correspond to objects that are just bright enough in the sub-mm wavelengths to be observed by ALMA, with albedos of 0.01 and 0.95, respectively.

Fig. 2 - Two configurations of SIM baselines to measure the shape of a projected elliptical disk discussed in the text. The elliptical disk represents a projected disk of a triaxial ellipsoidal shape, and the three dark solid lines are the three baseline orientations. In configuration 1 (left panel), the two baselines $B_1$ and $B_2$ are perpendicular to each other, and the third baseline $B_3$ aligns with the bisect between $B_1$ and $B_2$. In configuration 2 (right panel), the three baselines separate by 60º. In both case, the three baselines are ordered counterclockwise, and angle *φ* is measured counter-clockwise from the first baseline to the long-axis of the elliptical disk.

Fig. 3 - The two viewing geometries and rotation of the Haumea models to simulate the photometric light curves, visibility amplitude light curves, and visibility phase light curves from numerical simulations (Figs. 4, 5, and 6). Haumea has axial ratios of 1:0.86:0.54, a bright surface with a Minnaert parameter *k* of 0.85, and an albedo feature with about 40º in radius located at 45º longitude and having an albedo of 87% of the rest of the surface (Lacerda et al., 2008). Panel a has a sub-observer latitude 10º, and panel b 75º. The two panels show the rotation of the object (rotational phase noted under each sub-panel) as viewed from two viewing angles. The two thick lines, 20º and 110º from horizontal direction, in the second sub-panel in panel b mark the baseline orientations through which visibilities are calculated.

Fig. 4 - Photometric light curves (upper panels) and visibility amplitude light curves (lower panels) for near edge-on (left panels) and near pole-on (right panels) viewing angles as shown in Fig. 3. Solid lines are measured along a baseline 110º from horizontal direction, and dashed lines 20º from horizontal direction. The thick lines are for a uniform surface in the Haumea model, and the thin (gray) lines are for a surface with the dark-red-spot (DRS) in the same model.

Fig. 5 - Panel A shows the tracks of photocenter derived from visibility phase shift for the near edge-on viewing geometry of the Haumea model without surface brightness features shown in

Fig. 3a. Panel B shows the tracks for the same Haumea model with the DRS (panel b). The crosses mark the rotational phase 0.0, 0.25, 0.50, and 0.75. The diamonds are the rotational phases 30º and 60º between two crosses. The plots are tilted such that the two axes of the plot correspond to the baseline orientations through which the visibilities are measured.

Fig. 6 - Similar to Fig. 5, but for the near edge-on viewing angle of Haumea model shown in Fig. 3b. The thick line is for surface without brightness feature, and the thin line is for the surface with the DRS. The symbols mark the same rotational phases as in Fig. 5.

Fig. 7 - The visibility of a spherical object as a function of diameter measured in units of $\lambda/B_\perp$ for various Minnaert $k$. The solid line is for k=0.5, and other lines are for k=0.6 (dashed line), 0.8 (dash-dot), and 1.0 (dash-dot-dot-dot), respectively. SIM can measure the correlated flux to 1% uncertainty, and should be able to determine the limb-darkening parameter, $k$, to ±0.1 accuracy.

Fig. 8 - Limb-darkening correction factors for the visibility size of a spherical object. Symbols are from simulations, and lines are fit to an empirical function $C(\theta) = a_1 \exp(a_2 \theta) + a_3$, where $\theta$ is the size of object, and $a_1$, $a_2$, and $a_3$ are three constants for each Minnaert $k$. From the bottom line up, the solid lines correspond to Minnaert $k$ parameters from 0.3 to 1.0, with a step size of 0.05. When $k$=0.5, the disk has a uniform brightness, and the correction factor is unity. When $k$<0.5, the disk is limb-brightening; the measured visibility size overestimates the actual size. When $k$>0.5, the disk is limb-darkening; the measured size underestimates the actual size. The correction factors all converge to constants when the size of an object is approaches zero compared to fringe spacing.

**Fig. 1**

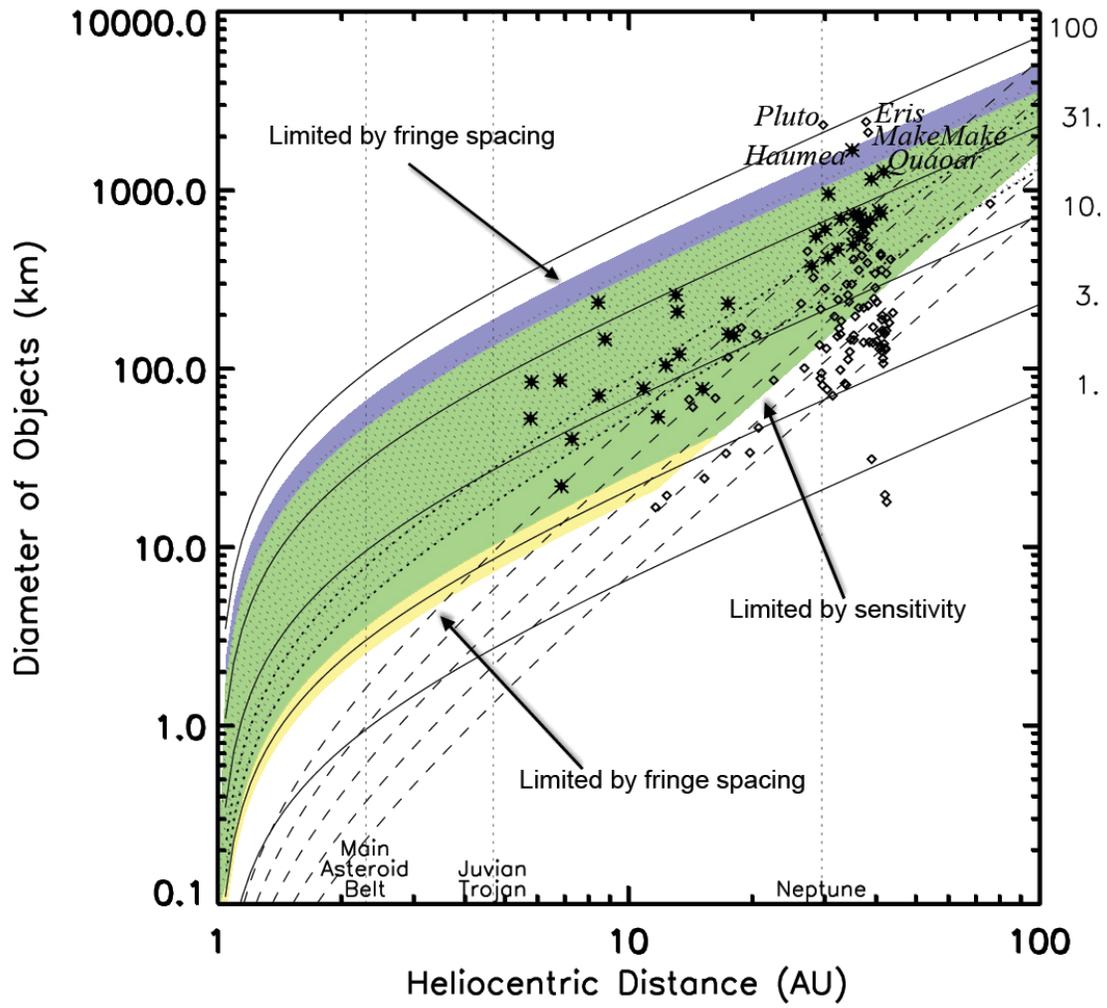

**Fig. 2**

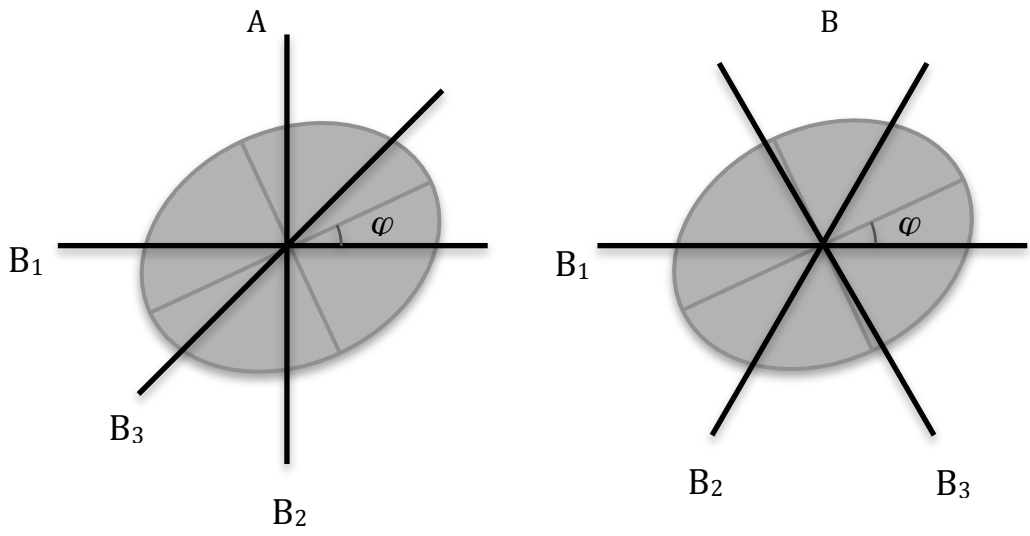

**Fig. 3**

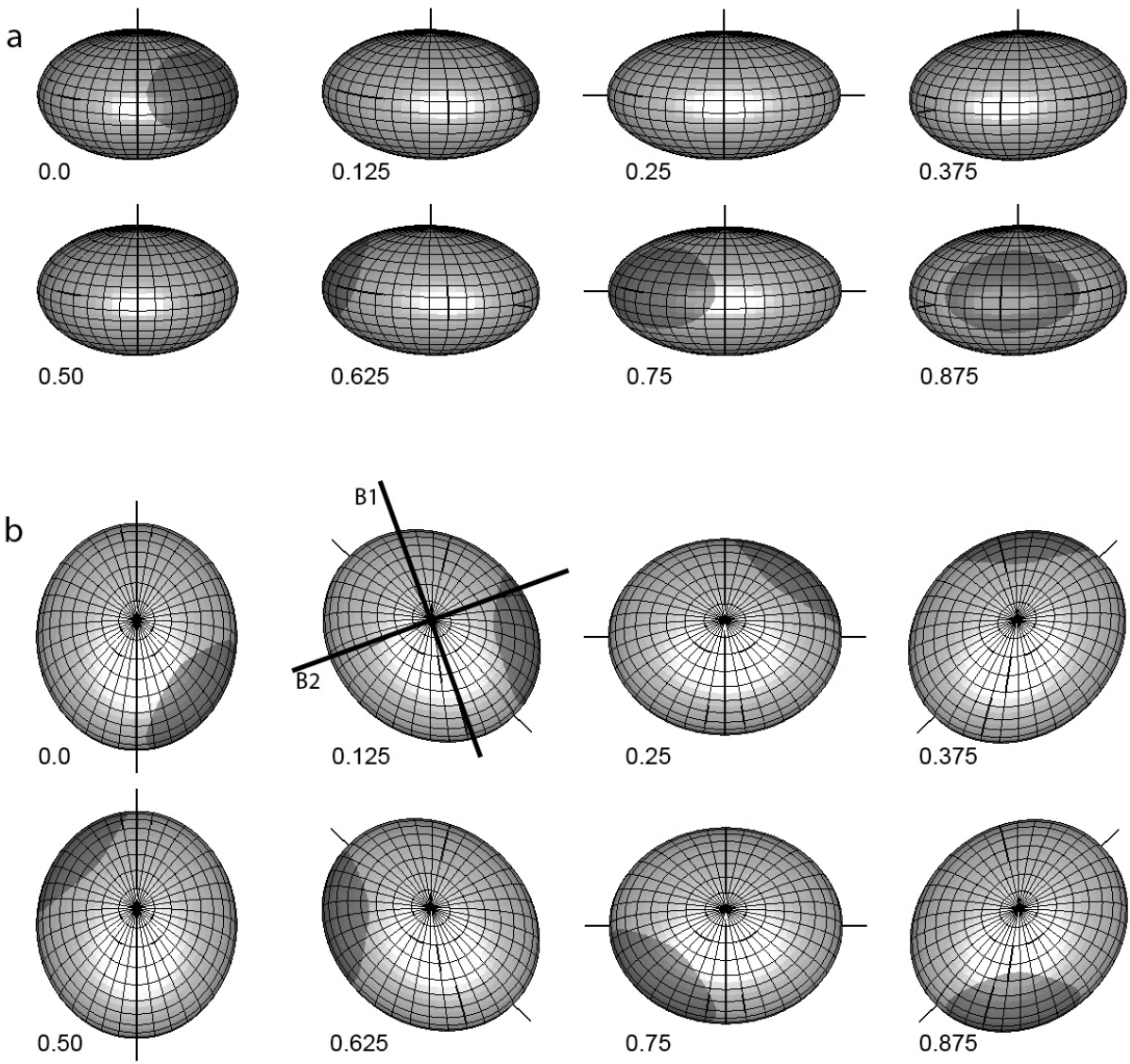

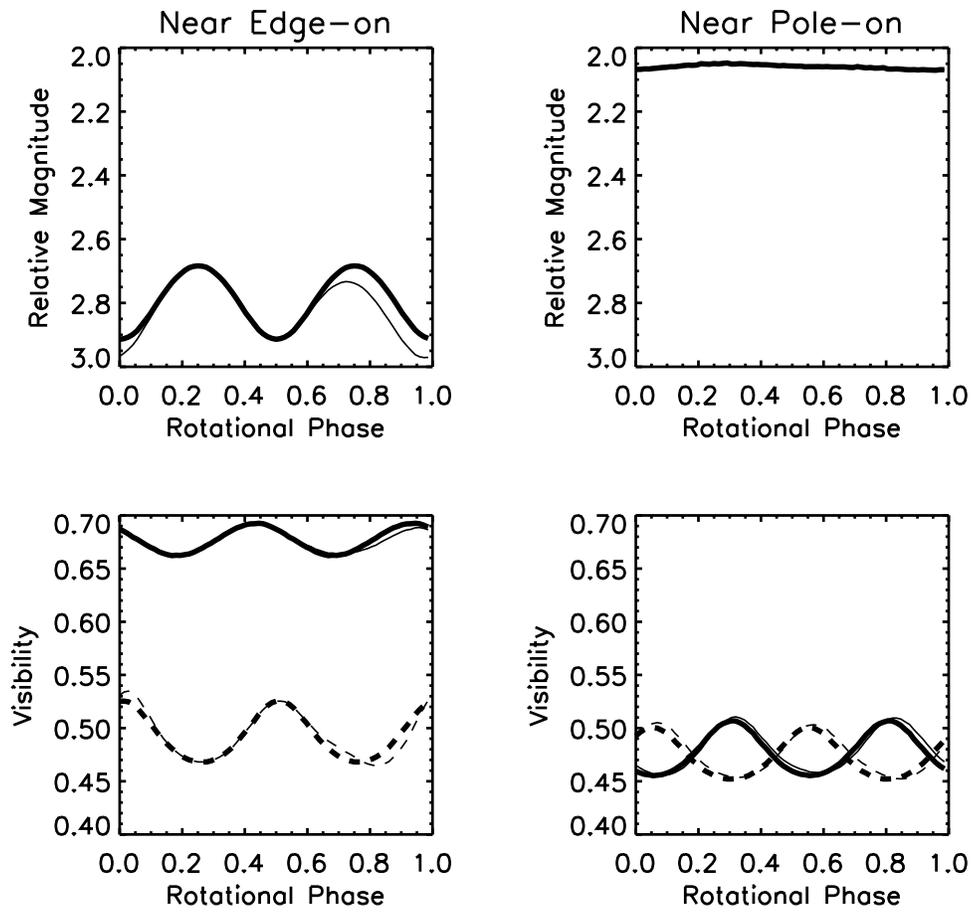

**Fig. 5**

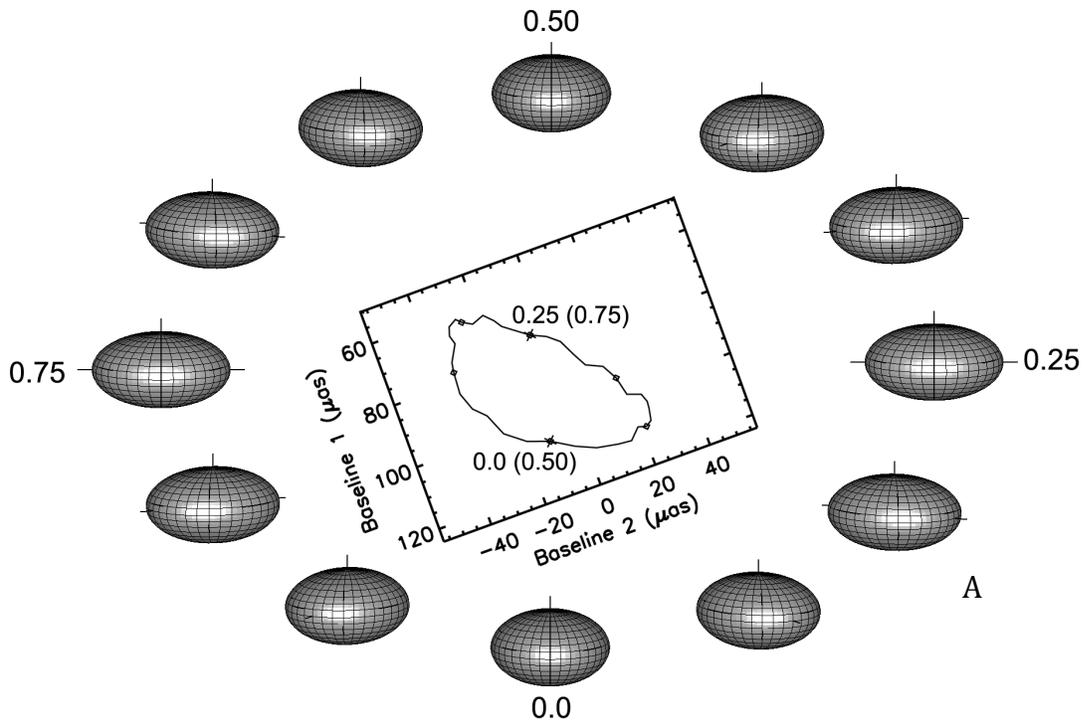

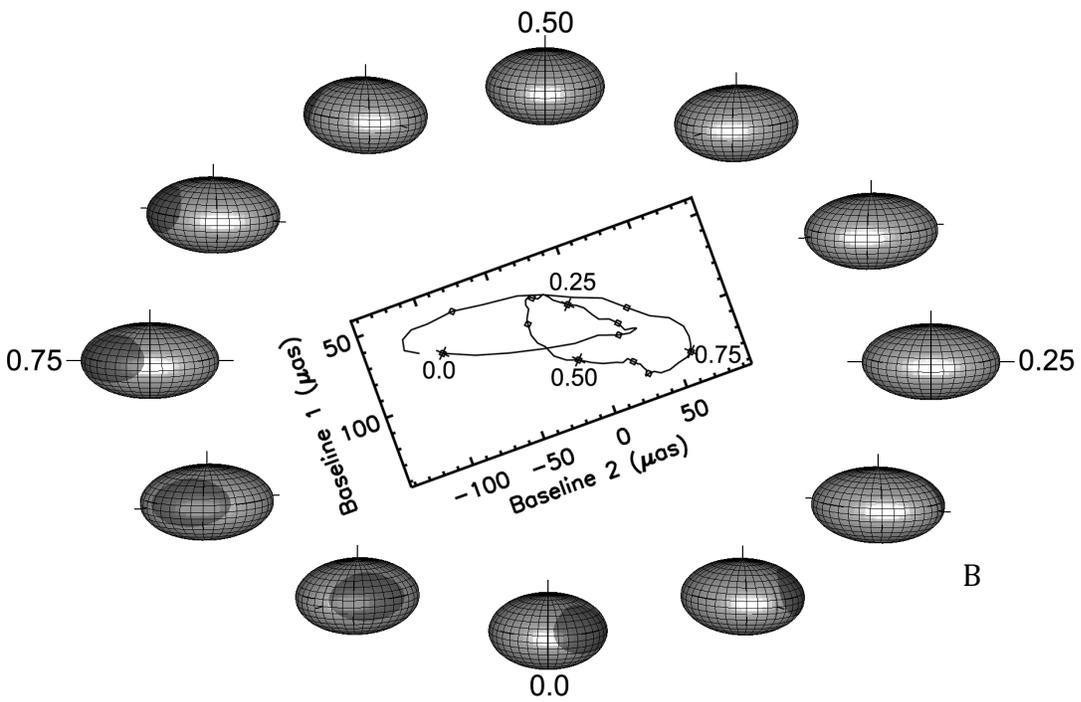

**Fig. 6**

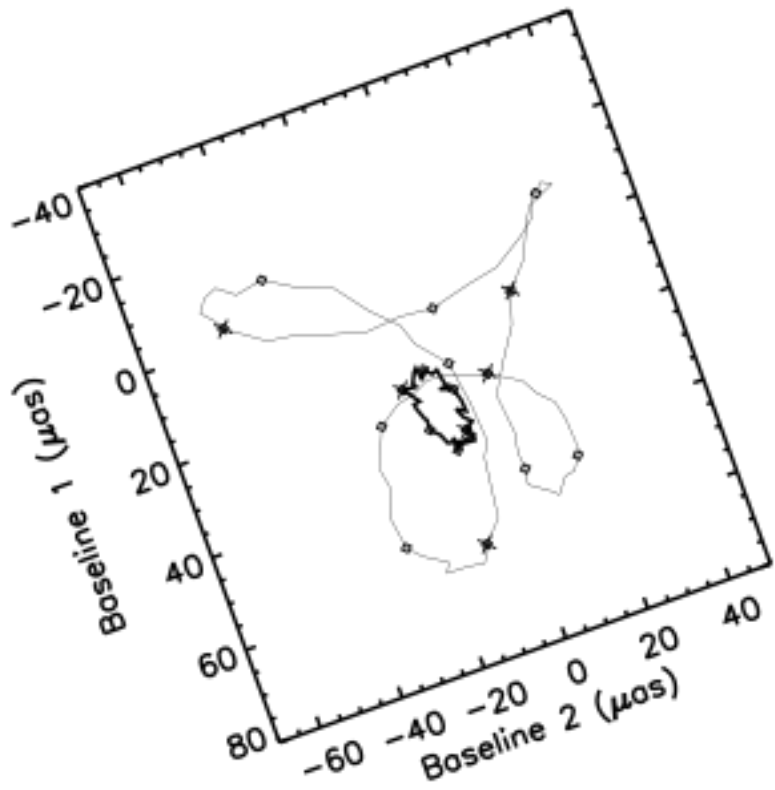

**Fig. 7**

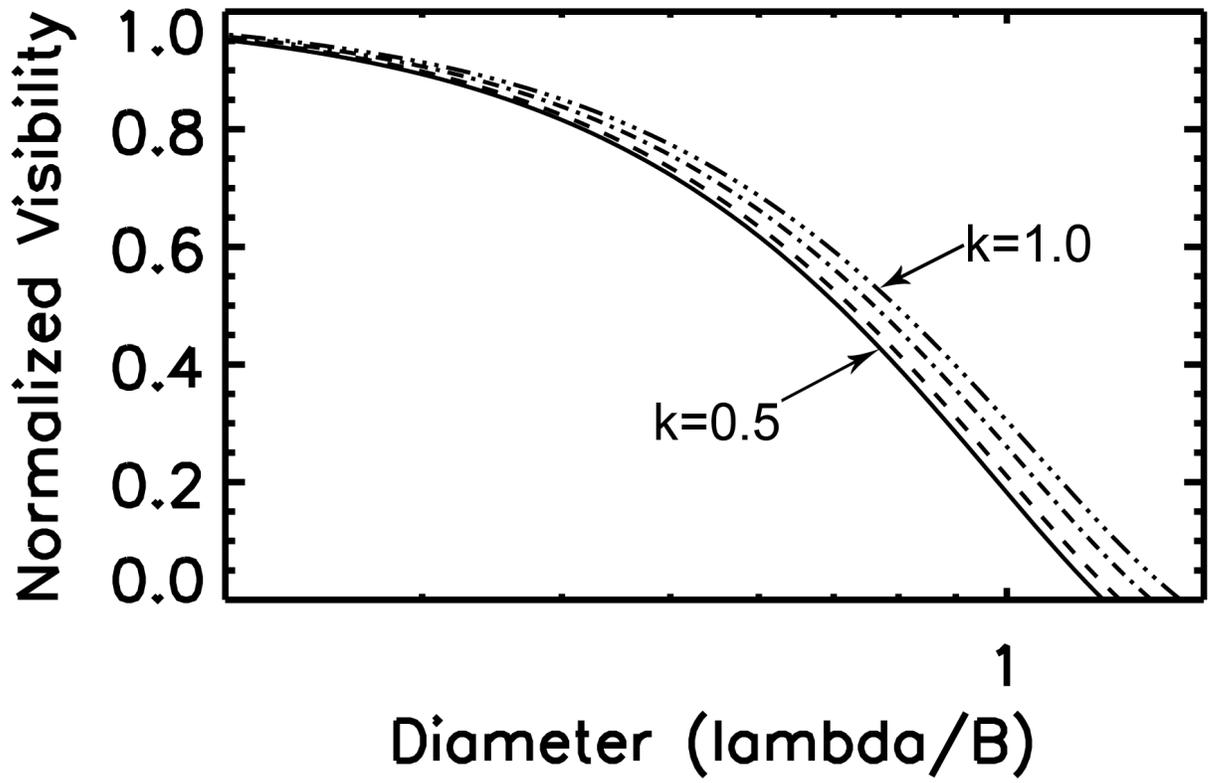

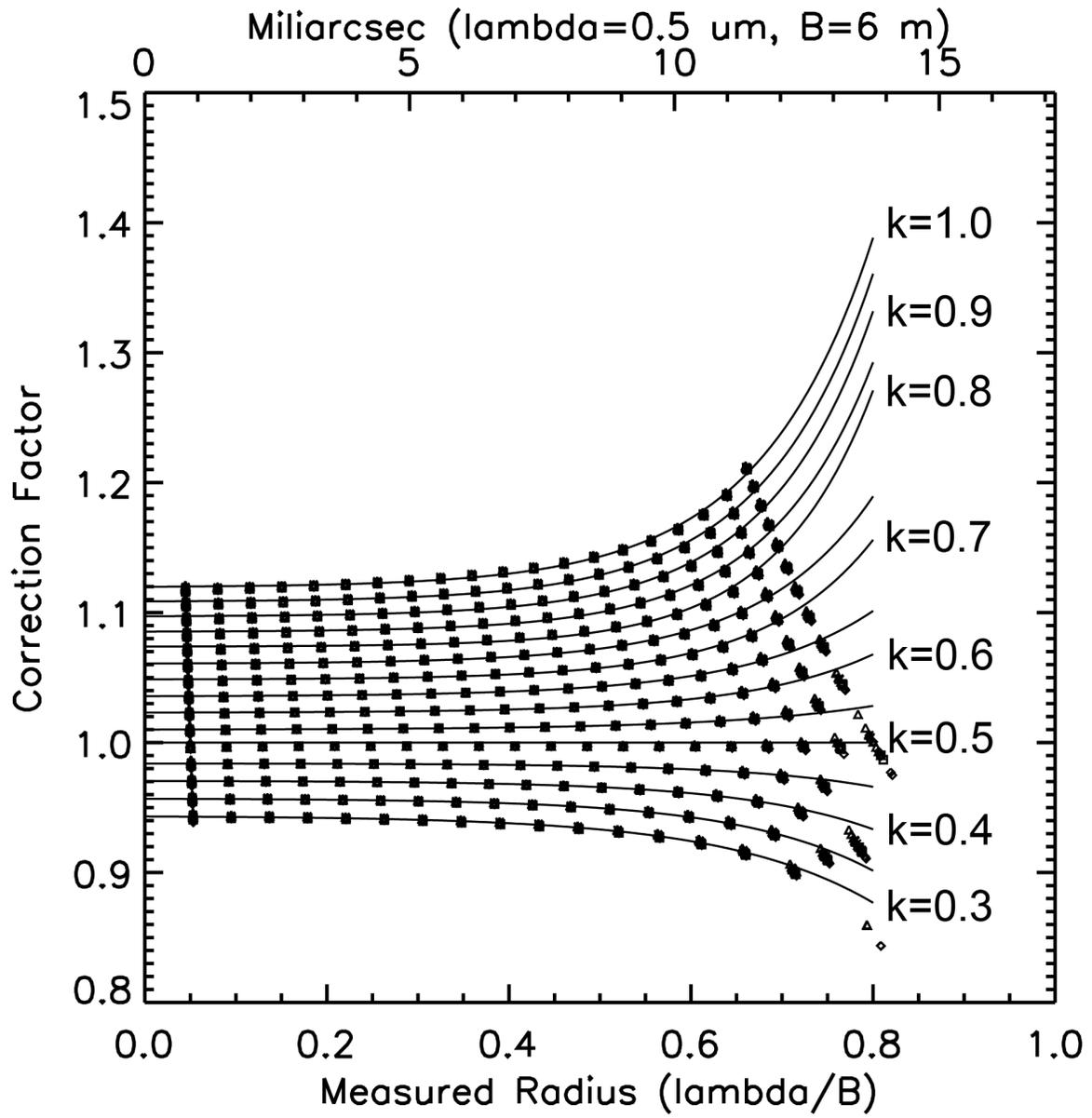